\documentclass[aps,nofootinbib,superscriptaddress,showpacs,preprintnumbers,  nofootinbibt]{revtex4-2}
\usepackage{amssymb}
\usepackage{amsmath}
\usepackage{amsthm}
\usepackage{graphicx}
\usepackage{alphalph,mathtools}
\usepackage{etoolbox}
\usepackage{color}
\usepackage{booktabs}
\usepackage{footnote}
\usepackage{makecell,tabularx}
\usepackage[leftcaption]{sidecap}
\usepackage{amssymb}
\usepackage{graphicx}
\usepackage{booktabs,dcolumn,caption}
\usepackage{rotating}
\usepackage{amsthm}
\usepackage{tikz}
\usepackage{caption}
\usepackage{subcaption}
\usepackage[most]{tcolorbox}
\usepackage{epsfig}
\usepackage{multirow}
\usepackage{eurosym}
\usepackage{dcolumn}
\usepackage{bm}
\usepackage{enumerate}
\usepackage{epstopdf}
\usepackage{amsmath}
\usepackage{bm}
\usepackage{amsfonts}
\usepackage{amssymb}
\usepackage{graphicx}
\usepackage{alphalph,mathtools}
\usepackage{etoolbox}
\usepackage{color}
\usepackage{booktabs}
\usepackage{footnote}
\usepackage{makecell,tabularx}
\usepackage{hyperref}

\usepackage{hyperref}
\hypersetup{colorlinks,citecolor=blue}
\hypersetup{colorlinks=true,linkcolor=red,filecolor=green,    urlcolor=blue}
\usepackage{setspace}
\def\be{\begin{equation}}
\def\ee{\end{equation}}
\def\bea{\begin{eqnarray}}
\def\eea{\end{eqnarray}}

\usepackage{orcidlink}
\begin{document}

\title{Probing Nonlinear Logarithmic Kalb-Ramond Black Holes: Particle Dynamics, Epicyclic Oscillations and Thermodynamic Signatures}
\author{Aftab Ansari~\orcidlink{0009-0009-2497-2166}}
\email[Email:]{aftab.student@ddugu.ac.in}
\affiliation{Mathematical Astrophysics Lab\\ Department of Mathematics and Statistics\\	Deen Dayal Upadhyaya Gorakhpur University, Gorakhpur, India.}

\author{Rajesh Kumar~\orcidlink{0000-0002-2582-7245}}
\email[Email:]{rajesh.mathstat@ddugu.ac.in, rkmath09@gmail.com}
\affiliation{Mathematical Astrophysics Lab\\ Department of Mathematics and Statistics\\	Deen Dayal Upadhyaya Gorakhpur University, Gorakhpur, India.}

\author{Praveen Kumar Dhankar~\orcidlink{0000-0002-8201-6019}}
\email[Email:]{praveen.dhankar@sitnagpur.siu.edu.in}
\affiliation{Symbiosis Institute of Technology, Nagpur Campus, Symbiosis International (Deemed University),  Pune-440008, India.}
\maketitle 
Looking beyond conventional linear and algebraic field coupling schemes, the present work investigated the nonlinear effects that become more significant in the strong-field regime of Kalb–Ramond (KR) black hole. By introducing coupling parameter $\beta$, we have investigated a modified class of KR black hole (BH) generated by a nonlinear logarithmic form, named as \textit{Logarithmic Kalb–Ramond black hole}. The logarithmic coupling introduces strong-field modifications to the spacetime geometry, leading to significant departures from the Schwarzschild and Reissner–Nordstrom like BHs. We numerically studied the horizon equation, which reveals the parameters-dependent two horizon structures- inner Cauchy and outer event horizons radii and showed that logarithmic deformation parameter $\beta$ significantly influences the horizon structure of BH. We analyze the motion of test particles using the effective potential formalism and derive the conserved energy and angular momentum for circular equatorial geodesics. The stability of circular orbits and the location of the innermost stable circular orbit (ISCO) are examined, revealing a dependence on the logarithmic coupling parameter $\beta$. The epicyclic frequencies (radial, vertical, and azimuthal) together with the associated periastron precession demonstrate that nonlinear logarithmic corrections can substantially modify quasi-periodic oscillation observables (QPO). We further investigate the Hawking temperature and energy emission rate, which show that the nonlinear coupling also impact a distinct imprints on the thermodynamic behavior and evaporation characteristics of the BH. Our results also identify the BH parameters as key regulators of the orbital dynamics, oscillatory properties, and thermal evolution, providing a unified framework for probing nonlinear KR gravity through strong-field astrophysical phenomena.

\maketitle


\maketitle
\tableofcontents
\section{Introduction}
\label{Intro}
The BHs represent one of the most remarkable predictions of General Relativity (GR) and provide an exceptional framework for probing the nature of gravity in the strong-field regime. Since its formulation by Einstein in 1915 \cite{1}, GR has successfully explained a wide range of astrophysical phenomena, spanning from the dynamics of the solar system to large-scale cosmological structures, and its predictions have been verified through numerous observational tests \cite{2}. Among the most intriguing consequences of GR are BHs-regions of spacetime where the gravitational field becomes so strong that even light cannot escape. The simplest BH solution is described by the Schwarzschild metric, which represents a static and spherically symmetric spacetime \cite{3}. Recent observational breakthroughs have significantly advanced our understanding of these objects and observational analyses indicate that both M87$^{*}$ and Sgr A$^{*}$ exhibit properties consistent with the theoretical concepts of BHs predicted by GR \cite{9,10,11,12,13,14,15,16,17,18,19}. In recent years, extensive studies of BHs and BH shadows  have further enriched our understanding of their observational signatures and optical characteristics \cite{21,22,23, 24,25,26,27,28,29,30,31,32,33}. Another important feature of test particle dynamics in BHs spacetimes is its close connection with quasi-periodic oscillations (QPOs), which are commonly detected in the X-ray emission of microquasars, BH binaries, and active galactic nuclei \cite{81, 84}. These oscillatory signals appear as distinct peaks in X-ray power density spectra and provide strong observational support for gravitational models operating in strong-field environments. Consequently, QPOs offer a valuable probe of matter dynamics in regions where spacetime curvature is extremely high. Since their discovery nearly three decades ago, they have attracted considerable attention in high-energy astrophysics because of their potential to reveal fundamental properties of compact objects. Observationally, QPOs are generally divided into two main categories: low-frequency QPOs, typically in the range of $0.1$-$30$ Hz, and high-frequency QPOs, which occur in the range of $40$-$450$ Hz \cite{HFQPOs and LFQPOs1,HFQPOs and LFQPOs2}. Despite notable progress in both theoretical modeling and observational analysis of BHs, the precise physical mechanism responsible for QPO generation is still not fully understood, which continues to motivate investigations within the frameworks of GR as well as modified theories of gravity \cite{85,86,87,88,89,90,91,92}. Because QPO frequencies are closely related to the characteristic orbital timescales near BHs, they contain essential information about the intrinsic properties of these compact objects, including their mass, spin, and the geometry of the surrounding spacetime \cite{82,83}. For this reason, QPO observations provide a powerful method for testing possible deviations from GR and constraining alternative gravitational models. In particular, within non-singular BH solutions such as hairy Bardeen-type geometries, QPO studies may place limits on exotic physical effects including nonlinear electrodynamics, scalar field contributions, or quantum gravity corrections that could alter the position of the ISCO or modify the epicyclic frequencies. Comprehensive discussions of the physical interpretation and observational properties of QPOs can be found in \cite{93}, while analyses across various BH geometries are presented in~ \cite{94,95,96,97,98,99,100,101}, with more recent developments reported in \cite{102,103}.
\par
Recent studies on Kalb-Ramond (KR) BHs motivated by string theory and Lorentz symmetry-violating gravity, have focused on both exact solutions and their observable signatures~\cite{intro1}. Several works have constructed  Schwarzschild-like BH solutions in KR gravity have been constructed by incorporating a non-minimally coupled antisymmetric tensor field whose vacuum expectation value induces spontaneous Lorentz violation and modifies the metric functions, horizon structure, and causal properties, with constraints obtained from Solar System tests and thermodynamics~\cite{Mukhopadhyaya2004, Kostelecky2009, SenGupta2001, RS2025, Paul2020}.
Extensions include electrically charged and AdS BH solutions, where the KR field modifies horizon structure and introduces new parameters affecting stability and causal structure~\cite{intro2, intro3, Shapiro2002, Hammond2002, Paul2019, Obukhov2014}. From a theoretical standpoint, non-rotating configurations are particularly important because they isolate the pure effects of the KR field without rotational complications, enabling detailed studies of thermodynamics, phase transitions, and stability, where Lorentz-violating parameters significantly modify Hawking temperature, entropy, and heat capacity~\cite{intro4, intro5}. Some recent works have emphasized observational aspects such as shadow radius, photon spheres, and weak deflection angles have been computed and constrained using observational data, showing that Lorentz-violating parameters significantly affect optical observables~\cite{intro6}. Moreover, some works extend the static geometries by embedding them in more realistic environments such as perfect fluid dark matter backgrounds or AdS spacetimes, revealing nontrivial effects on particle dynamics, QPOs, and accretion physics~\cite{intro7}. Additionally, X-ray timing missions such as NICER and NuSTAR have detected high-frequency QPOs from accretion disks around compact objects \cite{Ingram2019, Remillard2006}. These observations are sensitive to the BH metric through measurable features such as the BH shadow, orbital frequencies, and the radius of the ISCO \cite{Kato1998, Stella1999, Bambi2017, Shafee2006, PC2020}. Additionally, most existing analyses of BHs have been restricted to linear couplings, wherein the KR field is taken to be directly proportional to the spacetime curvature. In contrast, nonlinear couplings emerges from quantum corrections and higher-order, string-inspired interactions. These nonlinear contributions can significantly alter the near-horizon geometry, lead to modifications in BH thermodynamics, and affect test-particle dynamics. Consequently, even small deviations from the predictions of Schwarzschild or Reissner–Nordstrom (RN) solutions may provide observable signatures of underlying new physics~\cite{Bamba2016, Clifton2012}.

\par
In this work, we investigate the dynamics of test particles in the proposed \textit{Logarithmic KR BH model}, with particular emphasis on the role of the parameters charge $Q$, Lorentz-violating parameter $l$ and  logarithmic coupling $\beta$. Employing the effective potential formalism, we analyze the stability of circular orbits and derive analytical expressions for the conserved energy and angular momentum in terms of the BH parameters. The properties of the ISCO and the corresponding effective forces governing particle motion are examined in detail. Furthermore, we compute the epicyclic oscillations in the vicinity of the equatorial plane, obtaining explicit expressions for the radial, latitudinal and axial angular frequencies. We also explore the thermodynamic aspects of the model by evaluating the Hawking radiation temperature and the associated energy emission rate, and discuss the influence of the BH parameters on these quantities.
\par
The paper is structures as follows: a global introduction is presented in section~\ref{Intro}. In section~\ref{sec2a}, we discussed the previously investigated various KR BH models as the prerequisite of present work. In Sec.~\ref{sec2}, we introduce the nonlinear gravitational potential $f(r)$ for the logarithmic KR BH model and investigate its physical viability through an analysis of the curvature invariants and horizon structure. The admissible parameter space of the BH parameters is subsequently constrained numerically based on the existence of physically viable horizons. The next section~\ref{sec:particle-dynamics}, explores the particle dynamics around Logarithmic KR BH, covering (i) the effective potential in~\ref{subsec:veff},(ii) ISCOs in~\ref{subsec:isco} and (iii) the effective force in ~\ref{subsec:force}. Section~\ref{sec:HO} investigates harmonic oscillations as perturbations of circular orbits, analyzing (i) frequencies measured by a local observer~\ref{subsec:local-freq} and (ii) frequencies measured by a distant observer~\ref{subsec:distant-freq}. Subsequently, section~\ref{sec:EEHR} studied the emission energy  and Hawking radiation around Logarithmic KR BH model. The conclusions and summary of the work are discussed in section~\ref{sec:conclusion}.

\section{Various black hole solutions with the background of Kalb-Ramond field} \label{sec2a}
The Kalb--Ramond (KR) field, originally introduced in string theory as a rank-2 antisymmetric tensor field, provides a well-motivated extension of GR by incorporating higher-rank gauge degrees of freedom into the gravitational sector \cite{kalb1974classical, green1987superstring}. Owing to its intrinsic geometric nature, the KR field has attracted considerable attention in the study of modified gravitational theories and BH spacetimes. It naturally emerges in the low-energy effective action of string theory and has been shown to influence the structure, stability, and physical properties of compact objects through its antisymmetric field strength \cite{kao1996induced, Kostelecky2009, RS2025, Paul2020}.
\par

The KR field \(B_{ij}\) is a rank-two tensor field satisfying $B_{ij}=-B_{ji}$ and its associated field strength is described by the three-form $H_{ijk}=\partial_{[i}B_{jk]}$, where the square brackets denote antisymmetrization over the enclosed indices. The field strength remains invariant under the gauge transformation $ B_{ij}\rightarrow B_{ij}+\partial_{[i}\Gamma_{j]}$,  where $\Gamma_i$ is an arbitrary one-form. In analogy with the electric and magnetic decomposition of the electromagnetic field, the KR field can be decomposed into pseudo-electric and pseudo-magnetic components as  $ B_{ij}=\mathrm{E}_{[i}v_{j]}+\mathcal{E}_{ijk\delta}v^k\mathrm{B}^{\delta}$\cite{new1,new2}, subject to the orthogonality conditions $\mathrm{E}_iv^i=0=\mathrm{B}_iv^i$, where $v^i$ is a timelike four-vector and $\mathcal{E}_{ijk\delta}$ denotes the Levi-Civita tensor. Accordingly, $\mathrm{E}_i$ and $\mathrm{B}_i$ respectively are interpreted as the pseudo-electric and pseudo-magnetic components of the KR field interaction with in Maxwell electrodynamics.
 
 \par
 A general form action governing KR gravity, including the nonminimal coupling of the field to gravity, its self-interaction, and the coupling between the electromagnetic and, in the presence of a cosmological constant, can be expressed as~\cite{new1,new2,new5}, 

 \begin{equation}
    S = \frac{1}{2} \int d^4 x \sqrt{-g} \Big[ R- 2\Lambda + \frac{1}{6} H_{i j k} H^{i j k} - V + \xi_2 B^{i k} B^{j}_{k} R_{i j} +\xi_3 B^{i j} B_{i j} R  \Big] + \int d^4 x \sqrt{-g}~ L_{EM}
    \label{n1}
 \end{equation}
 Here, $\Lambda$ denotes the cosmological constant, while $\xi_2$ and $\xi_3$ are the nonminimal coupling constants describing the interaction between the gravitational field and the KR field. $V$ is the self-interaction potential of the KR field, defined by $V=V\left(B^{ij}B_{ij}\pm b^{ij}b_{ij}\right)$, where the KR field acquires a nonvanishing vacuum expectation value (VEV),
 $ \langle B_{ij}\rangle=b_{ij}$, thereby defining a background tensor field. The spontaneous breaking of local Lorentz symmetry is associated with this nonzero KR-field VEV induced by the self-interaction potential. The interaction between the electromagnetic and KR fields is described by the matter Lagrangian $L_{\rm EM} =-\frac{1}{2}\mathcal{F}^{ij}\mathcal{F}_{ij}
-\eta B^{ij}B^{mk}\mathcal{F}_{ij}\mathcal{F}_{mk}$,  where $\eta$ is the coupling contsnat, $\mathcal{F}_{ij}$ represents the field strength of the electromagnetic field and $8\pi G=1=c$.

\par
In this modified gravitational theory, a nonvanishing VEV of the KR field can act as a background tensor and induce spontaneous local Lorentz-symmetry breaking. The nonminimal couplings between the KR field, the curvature tensor and electromagnetic field,  modify the Einstein equations and consequently alter the geometry of BH spacetimes.\par
For a static and spherically symmetric spacetime, one generally considers the line element
\begin{equation}
ds^{2}
= -f(r)\,dt^{2}
+ \frac{dr^{2}}{f(r)}
+ r^{2}\left(d\theta^{2}+\sin^{2}\theta\,d\phi^{2}\right),
\label{metric}
\end{equation}
where $f(r)$ denotes the lapse function, which completely characterizes the gravitational field and determines the horizon structure as well as the causal properties of the BH spacetime.
\par
The different choices of the KR coupling and matter content lead to different forms of this lapse function. Some of the spherically symmetric BH solutions within the background of KR field  developed in the literature by using the action (\ref{n1}) are summarized below-
\par
(i)~ \textit{Schwarzschild-like black hole in KR field:} One of the simplest solutions is obtained by considering a nonminimally coupled KR field (with $\Lambda =0$, $L_{EM} =0$)~\cite{new3} 
\begin{equation}
  f(r)=  \frac{1}{1-l}-\frac{2M}{r}  
  \label{n2}
\end{equation}
where $M$ is the mass and $l$ is a dimensionless Lorentz-violating parameter associated with the KR VEV and its nonminimal coupling. In the limit $l \rightarrow 0$, the Schwarzschild BH is recovered from (\ref{n2}).
\par
(ii)~ \textit{Power-law hairy BH in background of KR field:} A more general static and spherically symmetric solution was obtained by Lessa et al.~\cite{new4} by considering a nonminimal coupling between the KR VEV and the Ricci tensor (with $\Lambda =0$, $L_{EM} =0$).
\begin{equation}
    f(r)=1-\frac{2M}{r}
+\frac{\Upsilon}{r^{\frac{2}{\lambda}}}
\label{n3}
\end{equation}
where $\lambda=|b|^2\xi_2$ is a dimensionless Lorentz-violating parameter determined by the magnitude of the KR VEV $b_{ij} b^{ij} =|b|^2 $ and $\Upsilon$ is an integration constant characterizing the KR-induced hair.

\par
(iii) ~For $\lambda=1$, Eq. (\ref{n3}) provides 
\begin{equation}
    f(r)=1-\frac{2M}{r}+\frac{\Upsilon}{r^2}
    \label{n4}
\end{equation}
which has the same radial dependence as the Reissner--Nordstrom (RN) solution. However, $\Upsilon$ is an independent quantity, representing a KR-induced Lorentz-violating hair~\cite{new4}, which is not directly  not be interpreted as an electric charge.
\par
(iv)~ \textit{Electrically charged KR BH:} The inclusion of an electromagnetic field ($L_{EM} \neq 0$) gives rise to a charged BH solution in KR configuration~\cite{new6, new7}.  Zheng-Qiao Duan et al. ~\cite{new5} obtained the charged KR BH with and without cosmological constant,

\begin{equation}
f(r)=
\begin{cases}
\dfrac{1}{1-l}-\dfrac{2M}{r}
+\dfrac{Q^2}{(1-l)^2r^2}
-\dfrac{\Lambda r^2}{3(1-l)}, & \Lambda\neq 0,\\[10pt]
\dfrac{1}{1-l}-\dfrac{2M}{r}
+\dfrac{Q^2}{(1-l)^2r^2}, & \Lambda=0.
\end{cases}
\label{n6}
\end{equation}
where $Q$ is the charge.\par
These solutions (i) -(iv) demonstrate that the presence of a KR background can modify the BH geometry through constant, power-law, charge-like, and cosmological term. The power-law solution(\ref{n3}-\ref{n4})  and the electrically charged solutions (\ref{n6}) show that KR-induced corrections can generate radial structures that closely resemble those of Schwarzschild-(A)dS Bhs or, RN, while possessing a different physical origin.
\par
 The subsequent studies also extended these results to  more general configurations such as gravitational lensing, BH shadows, particle dynamics, QPOs, etc, demonstrating that KR gravity admits a broad class of BH solutions with rich phenomenology~\cite{KR, Kalb-Ramond, gravityjunior2024lensing, kumar2020rotating, atamurotov2022particle, jumaniyozov2025pfdm}. Substantial progress has been made in constructing generalized KR BHs solutions by incorporating modified lapse function $f(r)$ that encode effective charge-like contributions, Lorentz-violating effects, and couplings to exotic matter fields. Such extensions give rise to significant departures from the standard Schwarzschild and RN geometries, often leading to nontrivial modifications of the horizon structure as well as the nature of spacetime singularities (see, e.g., Refs.~\cite{kr1, kr2, kr3, kr4} and references therein). 
\par
\section{Logarithmic KR Black Hole: A generalized nonlinear  ansatz of $f(r)$} \label{sec2}
Nevertheless, the majority of existing studies have been confined to linear and algebraic coupling schemes, wherein the KR field is assumed to couple proportionally to the spacetime curvature. From a more fundamental perspective, nonlinear coupling structures—particularly those of logarithmic or, exponential form, are well motivated by quantum corrections and higher-order interactions emerging in string-inspired gravitational theories \cite{Nojiri2011, Capozziello2011, Harko2014}. Such nonlinear effects are expected to play a crucial role in the strong-field regime, with potentially significant implications for near-horizon geometry, BH thermodynamics, and the dynamics of test particles~\cite{KR, Kalb-Ramond}. Despite these motivations, a systematic investigation of KR BHs incorporating nonlinear couplings, especially in conjunction with Lorentz-violating contributions, remains largely absent in the current literature.
\par
In the present investigation, a modified class of BH model discussed through a logarithmic coupling of the KR field with gravity. This framework generalizes existing models by incorporating two independent parameters: the Lorentz-violating parameter ($l$) representing deviations from local Lorentz symmetry, and the logarithmic coupling parameter ($\beta$) which encodes the nonlinear self-interaction of the KR field.\par 
By generalizing the algebraic structure of $f(r)$, we introduce a purely nonlinear logarithmic ansatz for the lapse function, characterized by the parameter $\beta$.

\begin{equation}
    f(r)
    =\frac{1}{1 - l \ln\!\left(1 + \frac{\beta M}{r}\right)}
- \frac{2M}{r}+ \frac{Q^2}{r^2 \left[1 - l \ln\!(1 + \frac{\beta M}{r})\right]^2}
 \label{f(r)}
\end{equation}
\par
It is to be noted here that \(f(r)\) in Eq.~(\ref{f(r)}) is constructed by introducing a nonlinear logarithmic modification to the  KR BH solution given in Eq.~(\ref{n6}) discussed by Zheng-Qiao Duan et al.~\cite{new5} (without cosmological constant), rather than deriving the solution independently from the action in Eq.~(\ref{n1}). There are also some more important rationale behind considering the logarithmic deformation in KR BH models, which are inadequately unexplored in the literature, such as- (i) the role of nonlinear KR couplings- logarithmic interactions, in shaping the horizon structure and governing geodesic stability;  (ii) the combined influence of the Lorentz-violating parameter $l$ and the deformation parameter $\beta$ on the effective potential and the location of the innermost stable circular orbit (ISCO).
\par

An important feature of the proposed metric function in Eq.~(\ref{f(r)}) is that it provides a unified framework encompassing several well-established BH solutions, such as it reduces to
\par
(a) Standard RN BH solution, $f(r)= 1- \frac{2 M}{r} + \frac{Q^2}{r^2}$ when $l=0$; \par (b) Schwarzschild BH solution; $ f(r)=1-\frac{2M}{r}$ for either $Q=0,l=0$ or $Q=0,\beta=0$; \par(c) Power-law hairy BH (\ref{n3}) and (\ref{n4}), when $l=0$ and charge is replace by KR-induced Lorentz violating hair $\Upsilon$; \par(d) Electrically charged BH (\ref{n6}) in the limiting position $r \rightarrow \frac{\beta M}{e-1}$.

\par
Thus, the proposed lapse function $f(r)$  does not represent an isolated deformation, but rather constitutes a generalized geometry that contains several known solutions within its parameter space. In the corresponding limits, the logarithmic and KR contributions become negligible or reduce to the appropriate form, can thereby recovered the expected results of the established BH models. This limiting behavior provides an important consistency check on the proposed construction (\ref{f(r)}) and, also enables a systematic investigation of how the nonlinear logarithmic KR coupling modifies the geometry and associated physical observables away from the well-known solutions.\par
\begin{figure}[t]
    \centering
 \includegraphics[width=\linewidth]{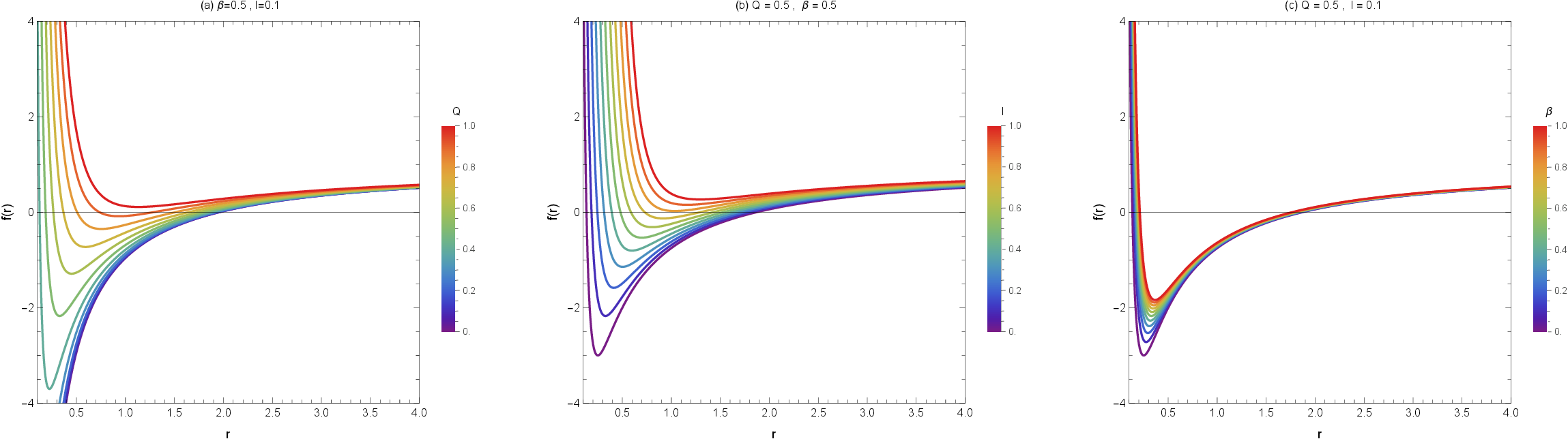}
    \caption{The variation of $f(r)$ for  Logarithmic KR BH within parameter-space ($0.0, 1.0$) of model parameters-$Q, l, \beta$.}
    \label{fig:fr_total}
\end{figure}

The graphical representations of function $f(r)$ are shown in figure~\ref{fig:fr_total}, which depict its variation within parameter space of $Q, l$ and $\beta$ (which are BH parameters in our model).  The observed variation of f(r) with larger values of $Q$, $l$ and $\beta$ further suggests a weakening of the effective gravitational attraction, consistent with the repulsive geometric contributions associated with these parameters. The effect of logarithmic coupling parameter $\beta$ on $f(r)$ appear primarily as higher strong gravity as comparison to  other BH parameters $Q$ and $l$. Consequently, Fig.~\ref{fig:fr_total} provides a comprehensive diagnostic of the interplay between charge effects, Lorentz-violating corrections, and nonlinear logarithmic coupling in determining the overall structure of the  BH model. From a physical perspective, the function $f(r)$ plays a central role, to determines the horizon structure, while its form also governs the behavior of timelike and null geodesics, effective potentials, perturbative dynamics and Energy Emission. 

\subsection{Curvature Invariants}\label{sec2.1}
The study of curvature invariants provides an effective way to examine the intrinsic geometry and regularity of the spacetime described by $f(r)$. In particular, quantities such as  the squared Ricci scalar  and the Kretschmann scalar are useful for characterizing the curvature structure and identifying possible singular behavior of the gravitational region.
\par
The Square of the Ricci scalar is defined by $R^{i j}R_{i j}$, which gives for the Logarithmic KR model (\ref{f(r)}) 
\begin{equation}
\begin{aligned}
R^{i j}R_{i j}
={}&\frac{1}{2r^{8}(\beta+r)^{4}\left[lL(r)-1\right]^{8}}
\Bigg\{
\Bigg[
2Q^{2}\!\left\{\beta^{2}(3l^{2}+3l+1)+2\beta(2l+1)r+r^{2}\right\}
+l^{2}L^{2}(r)\!\left\{2Q^{2}(\beta+r)^{2}-\beta^{2}lr^{2}\right\}
\\[-1pt]
&\quad
-2lL(r)\!\left\{
Q^{2}\!\left[\beta^{2}(3l+2)+4\beta(l+1)r+2r^{2}\right]
+\beta^{2}(l-1)lr^{2}\right\}
+\beta^{2}l(2l-1)r^{2}
\Bigg]^{2}
+4(\beta+r)^{2}\left[lL(r)-1\right]^{2}
\\[-1pt]
&\quad
\Bigg[
-l^{3}r^{2}(\beta+r)L^{3}(r)
+2l^{2}r^{2}(\beta+r)L^{2}(r)
-lL(r)\!\left\{r^{2}\!\left[\beta(l+1)+r\right]+Q^{2}(\beta+r)\right\}
+Q^{2}(\beta+2\beta l+r)+\beta lr^{2}
\Bigg]^{2}
\Bigg\}
\end{aligned}
\label{eq:K1}
\end{equation}
where,
\begin{equation}
L(r)=\log\!\left(1+\frac{\beta }{r}\right)
\end{equation}
The Kretschmann scalar is 
\begin{equation}
K = R_{h i j k}R^{h i j k}
\end{equation}
which gives for the Logarithmic KR model (\ref{f(r)})
\begin{equation}
\begin{aligned}
K={}&
\frac{1}{(\beta+r)^4}
\Bigg\{
\beta^2 l
\Big[
lL(r)\Big(lr^2L(r)+2(l-1)r^2-2Q^2\Big)
 +(2-6l)Q^2 +(1-2l)r^2
\Big]
-6Q^2(\beta+r)^2\big[lL(r)-1\big]^2\\
&\qquad
+12\beta lQ^2(\beta+r)\big[lL(r)-1\big] 
-2\beta lr^2(\beta+r)\big[lL(r)-1\big]^2
+4r(\beta+r)^2\big[lL(r)-1\big]^4
\Bigg\}^{2}
+ \\
& \quad
\frac{4\big[lL(r)-1\big]^2}{(\beta+r)^2}
\Bigg[
2\beta lQ^2 
-2Q^2(\beta+r)\big[lL(r)-1\big]
+\beta lr^2\big[1-lL(r)\big]
+2r(\beta+r)\big[lL(r)-1\big]^3
\Bigg]^2 
+ \\
&\quad 
\frac{4}{r^8\big[lL(r)-1\big]^4}
\Big[
lrL(r)\Big(-l(r+2)L(r)+r+4\Big)
+Q^2-2r
\Big]^2 .
\end{aligned}
\label{eq:K2}
\end{equation}

\begin{figure}[t]
    \centering
 \includegraphics[width=\linewidth]{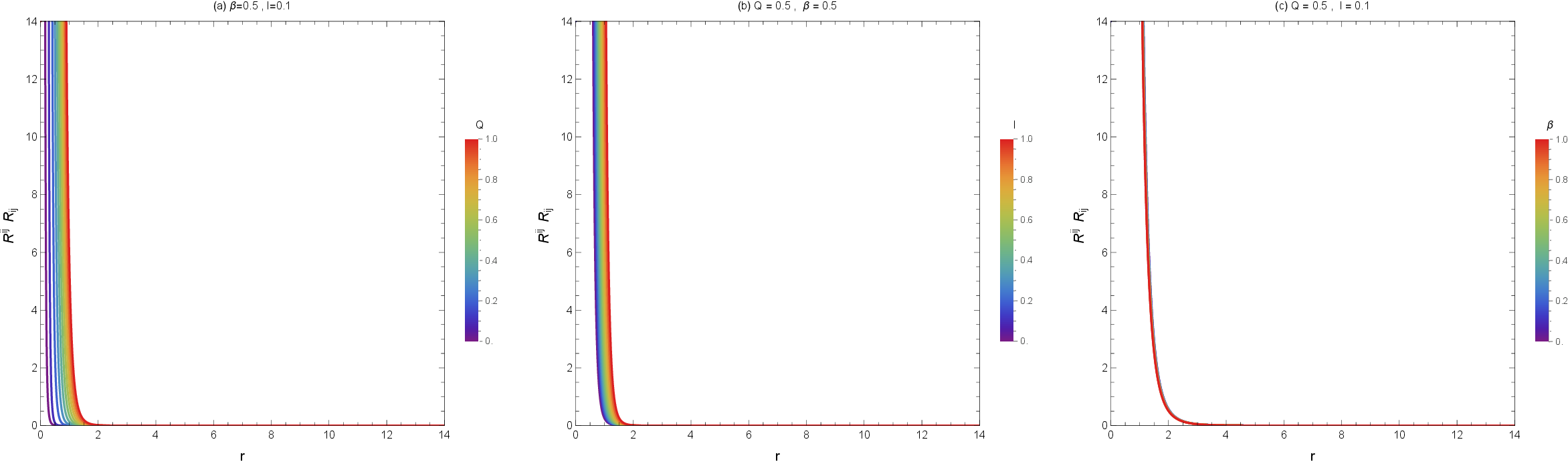}
    \caption{The variations of the Ricci Square   for Logarithmic KR BH within parameter-space ($0.0, 1.0$) of model parameters-$Q, l, \beta$.}
    \label{fig:Ricii2}
\end{figure}

\begin{figure}[t]
    \centering
 \includegraphics[width=\linewidth]{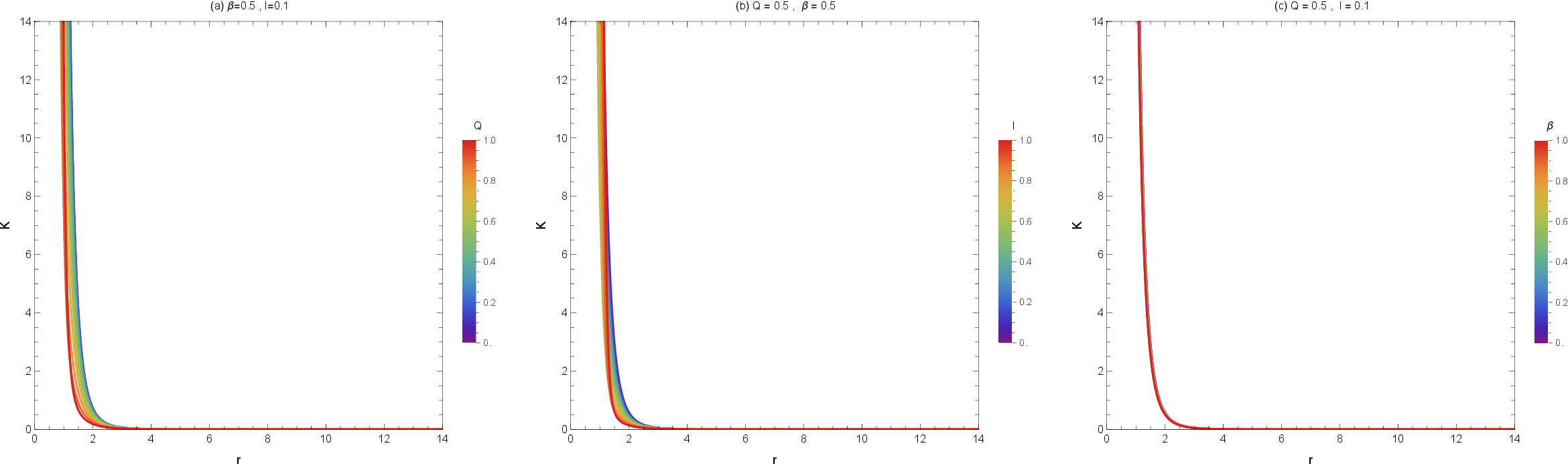}
    \caption{The variation of the Kretschmann Curvature for  Logarithmic KR BH within parameter-space ($0.0, 1.0$) of model parameters-$Q, l, \beta$.}
    \label{fig:Kreschmann}
\end{figure}

It can be straightforwardly verified from Eqs. (\ref{eq:K1}) and (\ref{eq:K2}) that Ricci square scalar and the Kretschmann scalar diverge near the center $r=0$ and also it is worth noticing that these curvatures approach zero as $r\rightarrow \infty$ along with $f(r)$ becoming constant (such behavior of curvature invariants also illustrated in figures- \ref{fig:fr_total}, \ref{fig:Ricii2} and \ref{fig:Kreschmann}). Therefore, spacetime (\ref{f(r)}) has a true singularity at $r=0$ and it becomes asymptotically flat as $r\rightarrow \infty$. On the BH horizon region dominated by logarithmic KR field, $f(r) \to 0$, implying that the proper time experienced by a local observer progresses markedly more slowly compared to the coordinate time, whereas in weak-field regimes or asymptotically flatness, the lapse function approaches unity, $f(r) \to 1 $, signifying the recovery of standard time evolution.
\subsection{Horizon structure of Logarithmic KR BH}\label{sec2.2}
 The horizon structure provides a direct characterization of the causal geometry associated with BH spacetime. The horizons of the spacetime are determined by
 \begin{equation}
     r_h = \{ r\in \mathbf{R} \quad |\quad  f(r) = 0 \}, \quad~\mathbf{R}~\text{is the set of real number}
 \end{equation}
 Because of the logarithmic dependence spacetime introduced by the KR deformation in eq.(\ref{f(r)}), the equation $f(r) = 0$ is transcendental and purely nonlinear and, in general, does not admit a closed-form analytical solution. We therefore determine the horizon radii numerically for some fixed values of the parameters $Q, l$ and $\beta$. The following tables-\ref{Table1}, \ref{Table2} and \ref{Table3},  describe the numerical roots of $f(r) = 0$ in the interval $r\in (0.0, 4.0)$ for specified range of $Q, l$ and $\beta$.
 \begin{table}[ht]
	\centering
	\setlength{\tabcolsep}{10pt}   
	\renewcommand{\arraystretch}{1.2} 
	\centering
	\caption{The numerical solution of $f(r) = 0$ in interval $r\in(0.0, 4.0)$ and for $Q\in (0.0, 1.0)$, $l=0.1, \beta= 0.5$}
	\label{tab:physical_properties}
	\begin{tabular}{@{}lccccc@{}}
		\hline
		\textbf{$Q$} & $r_{-}$(Cauchy inner horizon) & $r_{+}$(event horizon) & $\Delta r_h$ \\
		\midrule
		$0.0$  & - & 1.9544  & 1.9544\\
		$0.1$  & 0.0127139 & 1.94908 & 1.9363661 \\
		$0.2$  & 0.0380056 & 1.9328  & 1.8947949 \\
		$0.3$  & 0.0745506 & 1.90502  & 1.8304644\\
        $0.4$  & 0.123019 & 1.86459  & 1.741576 \\
         $0.5$  & 0.185192 & 1.8096  & 1.624408\\
          $0.6$  & 0.264403 & 1.73673  & 1.472327 \\
           $0.7$  & 0.366904 & 1.63978  & 1.272876\\
           $0.8$  & 0.50679 & 1.50473  & 0.99794 \\
           $0.9$  & 0.736527 & 1.27920 & 0.542673 \\
           $1.0$  & no solution & no solution & - \\
		\hline
	\end{tabular}
	\label{Table1}
\end{table}
\begin{table}[ht]
	\centering
	\setlength{\tabcolsep}{10pt}   
	\renewcommand{\arraystretch}{1.2} 
	\centering
	\caption{The numerical solution of $f(r) = 0$ in interval $r\in(0.0, 4.0)$ and for  $l \in (0.0, 1.0)$, $Q=0.5, \beta= 0.5$}
	\label{tab:physical_properties}
	\begin{tabular}{@{}lccccc@{}}
		\hline
		\textbf{$l$} & $r_{-}$(Cauchy inner horizon) & $r_{+}$(event horizon) & $\Delta r_h$ \\
		\midrule
		$0.0$  & 0.133975 & 1.86603  & 1.732056\\
		$0.1$  & 0.185192 & 1.8096 & 1.624408 \\
		$0.2$  & 0.245337 & 1.7489  & 1.503563 \\
		$0.3$  & 0.31406 & 1.68277  & 1.36871 \\
        $0.4$  & 0.392015 & 1.60937  &  1.21736\\
         $0.5$  & 0.481506 & 1.52558  & 1.044074 \\
          $0.6$  & 0.588305 & 1.42505  &  0.836745\\
           $0.7$  & 0.72981 & 1.28987  & 0.56006 \\
           $0.8$  & no solution & no solution  & -\\
           $0.9$  & no solution & no solution  & -\\
           $1.0$  & no solution & no solution  & -\\
           
		\hline
	\end{tabular}
	\label{Table2}
\end{table}
\begin{table}[ht]
	\centering
	\setlength{\tabcolsep}{10pt}   
	\renewcommand{\arraystretch}{1.2} 
	\centering
	\caption{The numerical solution of $f(r) = 0$ in interval $r\in(0.0, 4.0)$ and for $\beta \in (0.0, 1.0)$, $l=0.1, Q= 0.5$}
	\label{tab:physical_properties}
	\begin{tabular}{@{}lccccc@{}}
		\hline
		\textbf{$\beta$} & $r_{-}$(Cauchy inner horizon) & $r_{+}$(event horizon) & $\Delta r_h$ \\
		\midrule
		$0.0$  & 0.133975 & 1.86603  & 1.732055\\
		$0.1$  & 0.150733 & 1.85393 & 1.703197 \\
		$0.2$  & 0.162089 & 1.84227  & 1.680181 \\
		$0.3$  & 0.171046 & 1.83101  & 1.659964 \\
        $0.4$  & 0.178592 & 1.82013  & 1.641538 \\
         $0.5$  & 0.185192 & 1.8096  & 1.624408 \\
          $0.6$  & 0.191108 & 1.79939 & 1.608282 \\
           $0.7$  & 0.1965 & 1.7895  & 1.59300 \\
           $0.8$  & 0.201477 & 1.77989  & 1.57841 \\
           $0.9$  & 0.206116 & 1.77055  & 1.564434 \\
           $1.0$  & 0.210474 & 1.76147  & 1.550996 \\
		\hline
	\end{tabular}
	\label{Table3}
\end{table}
 For the parameter configurations considered in our numerical analysis, the equation $f(r)=0$ admits two distinct positive roots $0<r_{-} < r_{+}$, where following the standard interpretation of the horizon structure of logarithmic KR BH, we identify $r_{-} = r_{h}^{min}$ as the inner (Cauchy) horizon and larger root $r_{+} = r^{max}_{h}$ called, the outer (event) horizon, as illustrated in tables-\ref{Table1}, \ref{Table2} and \ref{Table3}.
 \par
 The existence of two distinct positive radii indicates that the logarithmic KR BH possesses a two-horizon structure analogous, to that encountered in other KR BH geometries~\cite{new2, new5}. The outer horizon $r_+$ constitutes the boundary of the BH region for an asymptotic observer whereas  the inner horizon $r_-$ is a Cauchy horizon separating the region between the two horizons from the innermost region. Accordingly, the radial structure can schematically be organized as
$$ 0< r_{-} < r< r_{+} $$
where these positive values mark changes in the causal character of the coordinates associated with the spacetime. The region $r>r_+$ corresponds to the exterior spacetime accessible to distant observers, while $r_-<r<r_+$ represents the intermediate region bounded by the inner and outer horizons.
\par
Importantly, it is to be noted that in the limit $Q=0$, the logarithmic KR BH admits only a horizon radius $r_{+}$, corresponding to the outer event horizon (while the inner horizon $r_{-}$ disappears), yielding a one-horizon structure analogous to the Schwarzschild like BHs (as can be seen in table-\ref{Table1}).  The numerical analysis of the horizon equation $f(r)=0$ imposes nontrivial constraints on the BH parameter space. In particular, no positive real horizon is found for $Q=1.0$ and also, the horizon solutions disappear at $l=0.8, 0.9,1.0$ which are also depicted in fig.\ref{fig:fr_total}(a, b). In contrast,  the positive horizon exists throughout the considered range of $\beta \in (0.0,1.0)$ (as can be seen in fig.\ref{fig:fr_total}(c)). Unlike $Q$ and $l$, whose larger values form the horizonless regime, logarithmic deformation parameter $\beta$ preserves the existence of a horizon. This analysis exhibits that $\beta$ plays a significant role in regulating the horizon structure of the KR BH model and provides a genuine deformation degree of freedom that controls the admissible BH sector which can be constrained by the requirement of horizon formation.

\par
An important feature of the present model is that the locations of both horizons depend on the KR BH parameters ($Q, l, \beta$). The numerical results shown in the corresponding plots of fig.\ref{fig:fr_total} also demonstrate this parameter dependence explicitly. Consequently, variation in these parameters modify the positions of $r_-$ and $r_+$, thereby altering the radial extent of the intermediate region and, more generally, the causal structure of the spacetime. The separation
\begin{equation}
    \Delta r_h = r_{+}  -r_{-}
\end{equation}
provides a useful measure of the departure from the degenerate-horizon configuration. The horizon analysis further shows that $\Delta r_h$, decreases systematically with increasing values of $Q, l$ and $\beta$, which indicates that the inner and outer horizons progressively approach each other as the strength of the charge, KR coupling, or logarithmic deformation (as depicted in tables-\ref{Table1}, \ref{Table2} and \ref{Table3}). The BH configuration moves toward a degenerate-horizon, or extremal state, characterized by $\Delta r_h\rightarrow 0$. Thus, the monotonic reduction of $\Delta r_h$ provides a useful measure of how these parameters drive the model toward extremal BH.
\subsubsection{Parameter-space constraint for logarithmic KR BH admissibility}
The parameter values adopted in the subsequent analysis of work are chosen on the basis of the horizon-existence constraints obtained from the numerical study of f(r)= 0 . Specifically, the ranges of $Q, l$ and $\beta$ are restricted to the parameter space in which physically admissible horizon solutions exist, thereby ensuring that the configurations analyzed throughout the present investigation correspond to BH spacetimes rather than horizonless geometries. Within this allowed parameter space, representative values are chosen to systematically examine the effects of the parameters $Q, l$ and $\beta$ on the dynamical and thermodynamic properties of the logarithmic KR BH.
\section{Particle dynamics around the Logarithmic KR BH}
\label{sec:particle-dynamics}
The motion of a test particle in the spacetime background of a Logarithmic KR BH can be conveniently analyzed using the Hamiltonian formalism. The Hamiltonian governing the particle dynamics is given by~\cite{e1}
\begin{equation}
    H = \frac{1}{2} g_{i j} p^{i} p^{j} + \frac{1}{2}\mu^{2},
    \label{Hamiltonian}
\end{equation}
where $\mu$ denotes the mass of a test particle and $p^{i}$ are the four-momentum, defined as $p^{i} = \mu u^{i}$, with  
$u^{i} = \frac{dx^{i}}{d\tau}$ representing the four-velocity and $\tau$ being the proper time of particle. \par
Within the Hamiltonian formalism, the equations governing the motion of the system can be expressed as follows-
\begin{equation}
    \frac{dx^{i}}{d\zeta} = \frac{\partial H}{\partial p_{i}}, 
    \qquad
    \frac{dp_{i}}{d\zeta} = -\frac{\partial H}{\partial x^{i}},
\end{equation}
where $\zeta=\tau/\mu$ is an affine parameter.
\par
Owing to the spacetime symmetries of the BH geometry, two conserved quantities naturally arise: the specific energy  $\mathcal{E}$ and the specific angular momentum $\mathcal{L}$, are expressible as 

\begin{equation}
\frac{p_t}{m} = \frac{1}{1 - l \ln\!\left(1 + \frac{\beta M}{r}\right)}
- \frac{2M}{r}+ \frac{Q^2}{r^2 \left(1 - l \ln\!\left(1 + \frac{\beta M}{r}\right)\right)^2} = -\mathcal{E}
\end{equation}

\begin{equation}
\frac{p_\phi}{m} = r^{2} \sin^{2}\theta \, \frac{d\phi}{d\tau} 
= \mathcal{L}
\end{equation}
Here, $\mathcal{E}$ and $\mathcal{L}$ represent the conserved specific energy and specific angular momentum of the particle, respectively and $m$ is the rest mass of the test particle.
\par
The components of the four-velocity $u^{i}$ (namely the time component $u^{t}$, the azimuthal component $u^{\phi}$ and the radial component $u^{r}$) govern the particle’s motion and satisfy the corresponding equations of motion derived from the spacetime geometry. Using these conserved quantities, the equations of motion reduce to 
\begin{align}
\dot{t}
&=
\frac{\mathcal{E}}
{\left(
\frac{1}{1-l\ln\!\left(1+\frac{\beta M}{r}\right)}
-\frac{2M}{r}
+\frac{Q^{2}}
{r^{2}\left(1-l\ln\!\left(1+\frac{\beta M}{r}\right)\right)^{2}}
\right)},
\label{eq8}
\\[2ex]
\dot{\phi}
&=
\frac{\mathcal{L}}
{r^{2}\sin^{2}\theta},
\label{eq9}
\\[2ex]
\dot{r}^{\,2}
&+
\left(
\epsilon
+
\frac{\mathcal{L}^{2}}
{r^{2}\sin^{2}\theta}
\right)
\left(
\frac{1}{1-l\ln\!\left(1+\frac{\beta M}{r}\right)}
-\frac{2M}{r}
+\frac{Q^{2}}
{r^{2}\left(1-l\ln\!\left(1+\frac{\beta M}{r}\right)\right)^{2}}
\right)
=
\mathcal{E}^{2}.
\label{eq10}
\end{align}
where  overdot ($.$) represents derivative with respect to proper time $\tau$.  Here, $\epsilon = 1$ characterizes to timelike particles, 
whereas $\epsilon = 0$ corresponds to null (lightlike) particles.
\\
Therefore, the Hamiltonian \eqref{Hamiltonian}, for the Logarithmic KR BHs model (\ref{f(r)}) reduces to the form,
\begin{equation}
    H = \frac{1}{2}
    \left(
        \frac{1}{1 - l \ln\!\left(1 + \frac{\beta M}{r}\right)}
- \frac{2M}{r}+ \frac{Q^2}{r^2 \left(1 - l \ln\!\left(1 + \frac{\beta M}{r}\right)\right)^2}
    \right)p_{r}^{2}
    + \frac{p_{\theta}^{2}}{2r^{2}}
    + \frac{\mu^{2}}{2f(r)}
        \left( V_{\text{eff}} - \mathcal{E}^{2} \right)
\end{equation}
where the effective potential $V_{\text{eff}}$ is given as
\begin{equation}
    V_{\text{eff}}(r,\theta)
    =
    \left( 1+\frac{\mathcal{L}^2 \csc ^2(\theta )}{r^2}\right) \left[ \frac{1}{1 - l \ln\!\left(1 + \frac{\beta M}{r}\right)}
- \frac{2M}{r}+ \frac{Q^2}{r^2 \left(1 - l \ln\!\left(1 + \frac{\beta M}{r}\right)\right)^2}\right]
    \label{Veff-general}
\end{equation}
In the analysis of particle motion, we restrict our attention to equatorial trajectories by considering the plane $\theta = \frac{\pi}{2}$, which guarantees that the motion can always be confined to a single plane without loss of generality. Consequently, the angular component of the effective potential reduces to $\frac{\mathcal{L}^{2}}{r^{2}}$, thereby simplifying the analysis and enabling a clearer characterization of the system’s dynamical properties.
\par
Since $\epsilon$ acts as a switching parameter between the time-like and null-like trajectories, its inclusion is fully consistent with the general geodesic formalism and facilitates a systematic investigation of particle dynamics. This parametrization provides a unified framework for describing both massive and massless particles within the same radial equation of motion. In present work, to account for geodesic motion, we have used $\epsilon = 1$ corresponds to timelike trajectories of massive particles. 

\subsection{Effective potential and Orbital dynamics} \label{subsec:veff}
The analysis of effective potential $V_{\rm eff}(r,\theta)$ plays a central role in understanding orbital dynamics of particles around BHs, as it encapsulates the combined influence of gravitational attraction and angular momentum barriers within a single framework. By reformulating the geodesic equations into a one-dimensional problem, the effective potential allows one to identify key features such as stable and unstable circular orbits, turning points, and the innermost stable circular orbit (ISCO)~\cite{Alomar2025, Ditta2025, Bouzenada2024,Saleem2025}. The nature of circular orbits is governed by the extrema of the effective potential- stable circular orbits correspond to its local minima, whereas unstable circular orbits are associated with its local maxima.
\par
\begin{figure}[t]
    \centering
    \includegraphics[width=\linewidth]{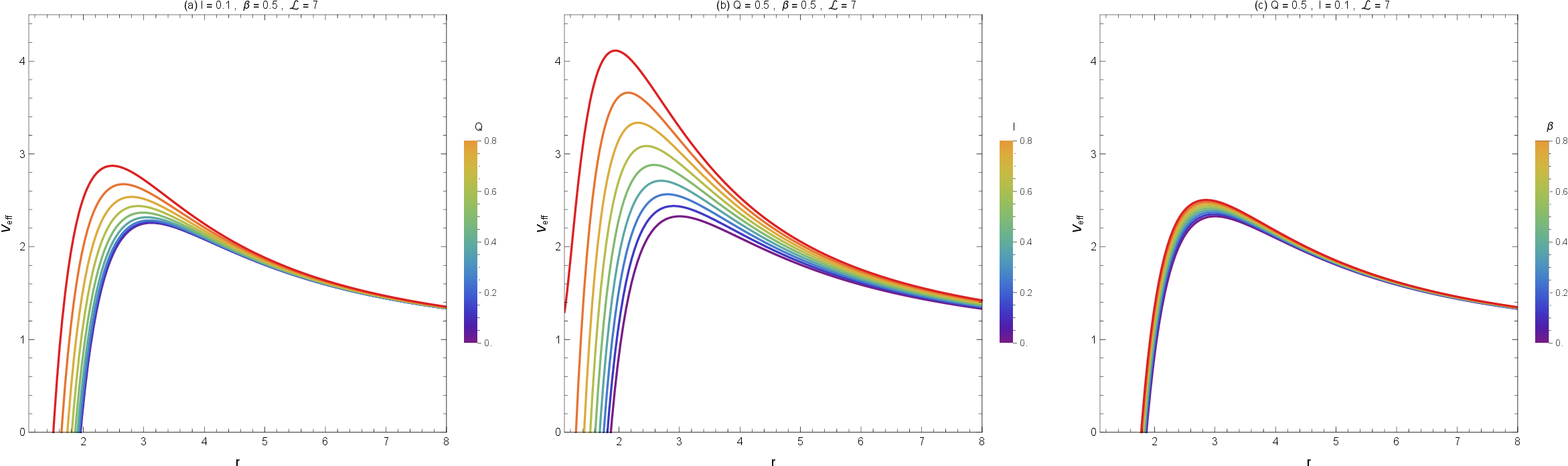}
    \caption{Effective potential $V_{\text{eff}}$ for the Logarithmic KR BH within parameter-space ($0.0, 0.8$) of  parameters-$Q, l, \beta$.}
    \label{fig:Veff}
\end{figure}

The variation of the effective potential $V_{\mathrm{eff}}$ with respect to the radial distance $r$ is illustrated in Fig.\ref{fig:Veff}. The extrema of the effective potential determine the nature of particle orbits, where minima correspond to stable circular orbits, while maxima represent unstable circular orbits. The presence of these extrema reflects the competition between gravitational attraction and centrifugal effects governing particle motion in the Logarithmic KR BH background. It is observed that increasing the parameters $Q$ and $l$ leads to an overall BH horizon as the charge parameter $Q$ increases, suggesting that stable circular orbits can exist at smaller radial distances compared to the Schwarzschild case. A similar trend is observed with increasing $l$, which further modifies the near-horizon structure of the spacetime. The variations in the parameter $\beta$ produce only a small change in the effective potential, indicating a comparatively weaker influence on orbital structure.
\par
From Fig.\ref{fig:Veff}, it is also evident that the height of the potential barrier increases with larger values of $Q$ and $l$, which implies that particles require greater energy to escape the gravitational well or transition between orbital regions. This behavior points to a stronger confinement of particle motion in the vicinity of the BH. Moreover, the inward shift of the potential minima signifies enhanced spacetime curvature near the horizon, which may have direct consequences for the location of the ISCO and the behavior of epicyclic oscillations.

Physically, these results demonstrate that the parameters $Q$ and $l$ play a significant role in shaping the orbital dynamics of test particles, while the parameter $\beta$ contributes only marginally. The modifications in the effective potential provide important insight into how the Logarithmic KR BH departs from the Schwarzschild geometry and may lead to observable effects in accretion dynamics and strong-field astrophysical phenomena.
\par
The circular orbits for the equatorial plane can be obtain by the following conditions~\cite{e3} 
\begin{equation}
V_{\text{eff}}(r)=\mathcal{E}^{2},
\qquad
\frac{dV_{\text{eff}}}{dr}=0
\label{condition}
\end{equation}

As shown in Fig.~\ref{fig:Veff}, increasing the parameter $Q$ deepens the potential and shifts the minima closer to the horizon and parameters $l$ and $\beta$ have comparatively weaker influence.

For solving the~\eqref{condition}, we obtain the circular orbits lead to the exact expressions for the specific energy and angular momentum

     \begin{equation}
\mathcal{E} = \sqrt{2} \, \sqrt{
\frac{
\left(
\frac{Q^2}{r^2 A^2(r)} + \frac{1}{A(r)} - \frac{2}{r}
\right)^2
}{
\frac{4 Q^2}{r^2 A^2(r)} + \frac{2}{A(r)} - \frac{6}{r} 
+ \frac{2 \beta l Q^2}{r^3 \left(\frac{\beta}{r} + 1\right) A^3(r)} 
+ \frac{\beta l}{r \left(\frac{\beta}{r} + 1\right) A^2(r)}
}
}
\label{E}
\end{equation}

\begin{equation}
\mathcal{L} = \sqrt{
\frac{
r^3 \left(
-\frac{2 \beta l Q^2}{r^4 \left(\frac{\beta}{r} + 1\right) A^3(r)}
-\frac{2 Q^2}{r^3 A^2(r)}
-\frac{\beta l}{r^2 \left(\frac{\beta}{r} + 1\right) A^2(r)}
+\frac{2}{r^2}
\right)
}{
\frac{4 Q^2}{r^2 A^2(r)} + \frac{2}{A(r)} - \frac{6}{r}
+ \frac{2 \beta l Q^2}{r^3 \left(\frac{\beta}{r} + 1\right) A^3(r)}
+ \frac{\beta l}{r \left(\frac{\beta}{r} + 1\right) A^2(r)}
}
}
\label{L}
\end{equation}
where $A(r) =1-l \log \left(\frac{\beta }{r}+1\right)$.
\begin{figure}[t]
    \centering
    \includegraphics[width=\linewidth]{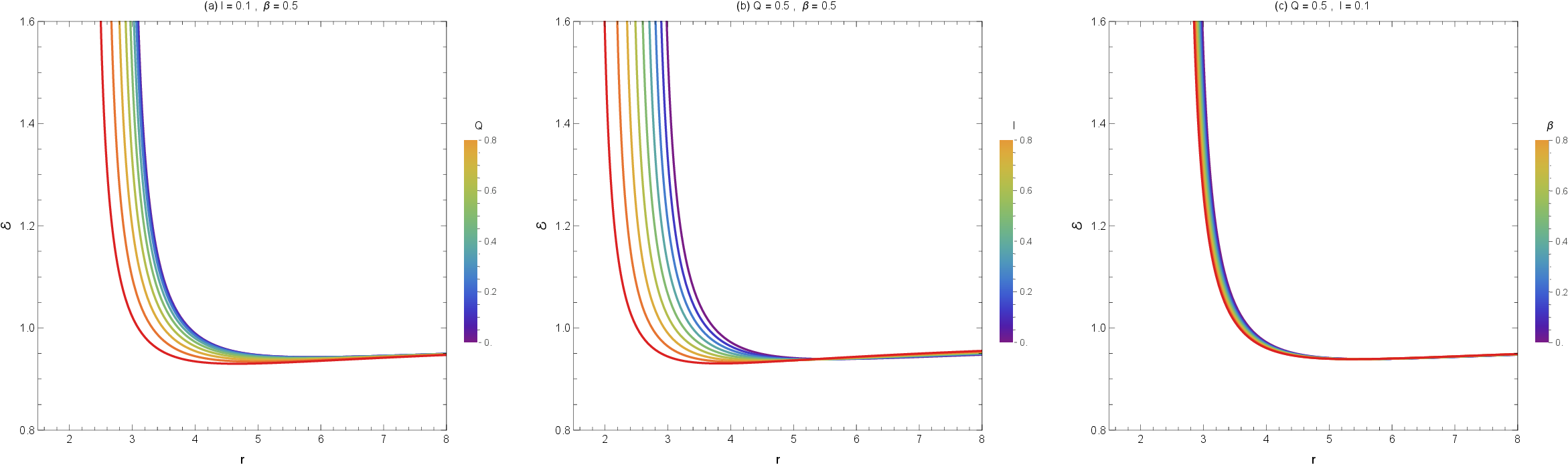}
    \caption{Specific energy $\mathcal{E}$ of circular orbits of the Logarithimic KR BH within parameter-space ($0.0, 0.8$) of  parameters-$Q, l, \beta$.}
    \label{fig:Eofr}
\end{figure}
\begin{figure}[t]
    \centering
    \includegraphics[width=\linewidth]{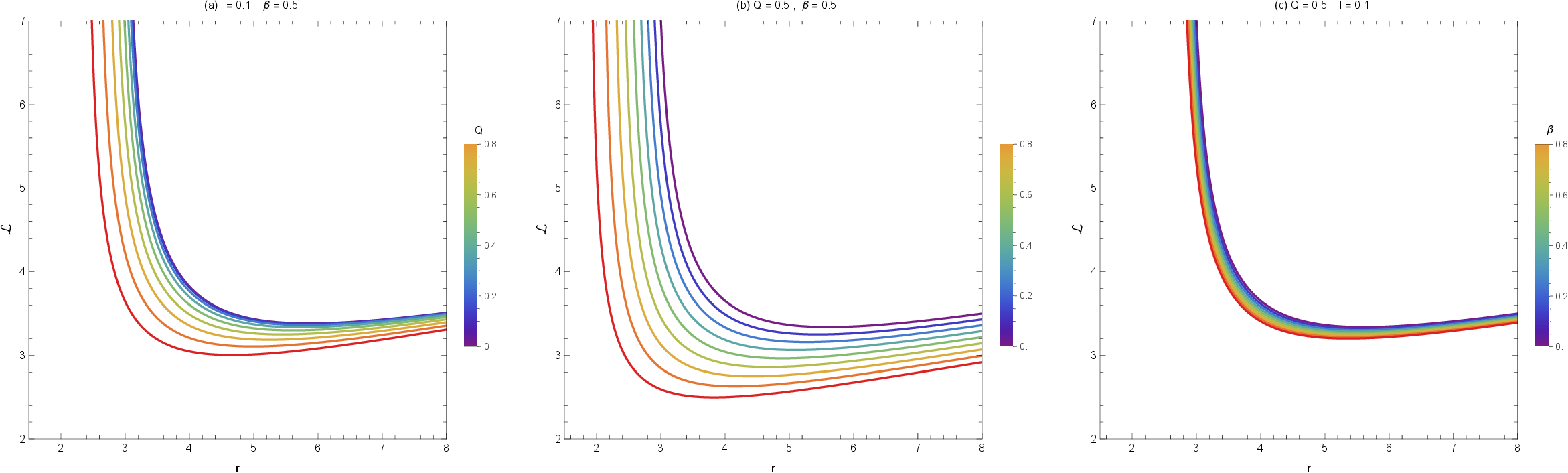}
    \caption{Specific angular momentum $\mathcal{L}$ of circular orbits of the Logarithmic KR BH within parameter-space ($0.0, 0.8$) of  parameters-$Q, l, \beta$.}
    \label{fig:Lofr}
\end{figure}
Figure~\ref{fig:Eofr} and~\ref{fig:Lofr} respectively illustrate the radial variation of the energy $\mathcal{E}$ and angular momentum $\mathcal{L}$ for circular equatorial orbits around the logarithmic KR BH. It is evident that $\mathcal{E}$ and $\mathcal{L}$ decrease with increasing values of the parameters $Q$ and $l$, while they exhibits a monotonic increase with the radial coordinate $r$. In contrast, the parameter $\beta$ has only a negligible effect on the overall energy profile and anugular momentum. From a physical standpoint, the reduction in energy indicates that particles require less binding energy to sustain stable circular motion, potentially enhancing orbital confinement in the vicinity of the BH. The reduction in angular momentum  further implies that particles require less rotational support to maintain circular motion, indicating an enhanced gravitational influence in the Logarithmic KR BH. The energy and angular momentum profiles also display a steep variation in the near horizon-region, followed by a gradual flattening at larger radii. Furthermore, the $\mathcal{E}$ and $\mathcal{L}$ curves leftward shift of the curves with increasing $Q$ and $l$ suggests that stable bound orbits can persist closer to the central object. These features imply that the parameters $Q$ and $l$ play a significant role in determining energetically favorable orbital configurations, with direct consequences for the ISCO structure, epicyclic oscillations, and associated observational signatures such as QPOs in accretion disks. Additionally, the comparatively higher particle energy relative to the Schwarzschild solution case highlights clear deviations from the classical BH scenario, underscoring the impact of logarithmic KR parameters on orbital dynamics.
\subsection{ Innermost stable circular orbit dynamics}\label{subsec:isco}
The dynamics of the Innermost Stable Circular Orbit (ISCO) plays a central role in understanding the motion of particles in strong gravitational fields. The ISCO represents the smallest radius at which a test particle can maintain a stable circular orbit; inside this radius, even small perturbations lead to an inevitable inspiral into the BHs. In the framework of GR, the ISCO depends sensitively on the spacetime geometry, including parameters such as mass, spin, and possible additional fields and Its location determines key astrophysical observables, including the inner edge of accretion disks, efficiency of energy extraction, and emitted radiation spectra. In modified gravity scenarios or in the presence of additional fields (KR), deviations in ISCO radius and associated orbital frequencies provide a powerful probe to test departures from standard BH solutions and to constrain underlying gravitational theories~\cite{Mustafa2025,Javed2025,Bouzenada2025}. Since, the stable and unstable circular orbits can be determined from the minima and maxima of the effective potential $V_{\rm eff}(r,\theta)$. However, when the effective potential depends on the particle’s angular momentum as well as additional physical parameters, this picture changes significantly and typically exhibits two extrema for a given angular momentum-corresponding to stable and unstable circular orbits. These two extrema merge only at a critical value of the angular momentum, signaling the onset of the ISCO.
\par
The ISCO radius $r=r_{\mathrm{ISCO}}$ can be characterized by the following,
  \begin{equation}
V_{\text{eff}}(r)=\mathcal{E}^{2},\qquad 
\frac{dV_{\text{eff}}}{dr}=0,\qquad
\frac{d^{2}V_{\text{eff}}}{dr^{2}}=0        
 \end{equation}
\begin{figure}[t]
    \centering
    \includegraphics[width=\linewidth]{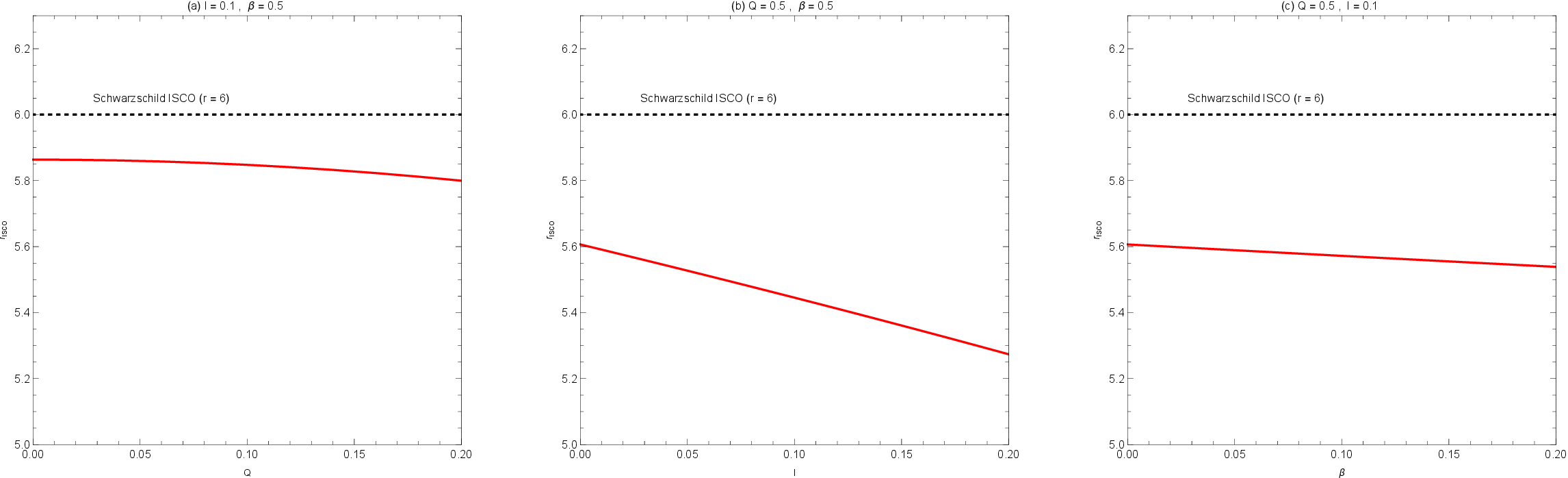}
    \caption{ISCO radius as a function of the parameters 
    $Q$, $l$ and $\beta$ for the Logarthimic KR BH.}
    \label{fig:ISCO}
\end{figure}

Figure~\ref{fig:ISCO} presents the behavior of the equatorial ISCO radius, $r_{\mathrm{ISCO}}$, for the Logarithmic KR BH. It is evident that increasing the parameter $l$ leads to a pronounced reduction in the ISCO radius [cf. Fig.~\ref{fig:ISCO}(b)]. In contrast, variations in $Q$ and $\beta$ induce only small changes, resulting in an almost invariant ISCO radius, as illustrated in Fig.~\ref{fig:ISCO}(a,c). This clearly indicates that the parameter $l$ exerts a dominant influence on orbital stability compared to $Q$ and $\beta$. The monotonic decrease of $r_{\mathrm{ISCO}}$ with increasing $l$ signifies an inward shift of the stability boundary, reflecting modifications in the near-horizon geometry and implying that stable circular motion can be sustained deeper within the gravitational potential well. From a physical perspective, the reduction in the ISCO radius is of considerable importance, as it determines the inner edge of accretion disks and strongly affects accretion efficiency, emitted radiation, and orbital dynamics in the strong-field regime. A smaller ISCO radius corresponds to a more compact region of stable motion, potentially enhancing relativistic effects such as gravitational redshift and altering epicyclic frequencies. Furthermore, the inward displacement of the ISCO toward the event horizon may have direct implications for QPOs and the observable properties of accreting matter. In the limiting case $Q = l = \beta = 0$, the ISCO radius reduces to $r_{\mathrm{ISCO}} = 6$, consistent with the standard Schwarzschild solution~\cite{ISCO} (as display the black-dotted line in fig. ~\ref{fig:ISCO}). Notably, the ISCO radii for the logarithmic KR BH are systematically smaller than those of the Schwarzschild BH model, indicating that stable circular orbits can exist closer to the central object.

\subsection{Effective force analysis} \label{subsec:force}
The effective force analysis of particle motion around BHs provides an intuitive framework to interpret the underlying geodesic dynamics by recasting the radial equation of motion in terms of an equivalent force derived from the gradient of the Effective potential $V_{\rm eff}$. In this approach, the competing influences of gravitational attraction, centrifugal repulsion, and possible additional fields are encoded in a single function, allowing one to identify stable and unstable circular orbits through extrema conditions~\cite{e2}. Such analysis is particularly useful for examining the impact of KR fields on orbital stability, capture cross-sections, and accretion processes. In this work, we analyze particle motion in the spacetime of an Logarithmic KR BH, where the modified gravitational framework allows for the emergence of both attractive and repulsive gravitational force regimes. The behavior of the effective force near the horizon and at large distances further provides insight into strong-field deviations from classical BH solutions and serves as a diagnostic tool for testing alternative theories of gravity~\cite{e3, e4}.\par
The effective force experienced by the particle is derived from Eq.~\eqref{Veff-general} and, is defined by
\begin{equation} 
    F = -\frac{1}{2}\frac{dV_{\text{eff}}}{dr},
\end{equation}
which yields:
\begin{equation}
  F =\frac{1}{2 r^5} \Bigg[\frac{2 \beta  l Q^2 \left(\mathcal{L}^2+r^2\right)}{(\beta +r) A^3(r)}+\frac{\mathcal{L}^2 \left(\beta  l r^2+4 Q^2 (\beta+r)\right)+\beta  l r^4+2 Q^2 r^2 (\beta +r)}{(\beta +r)A^2(r)}+\frac{2 \mathcal{L}^2 r^2}{A(r)}-2 r \left(3 \mathcal{L}^2+r^2\right) \Bigg]
\end{equation}
\begin{figure}[t]
        \includegraphics[width=\linewidth]{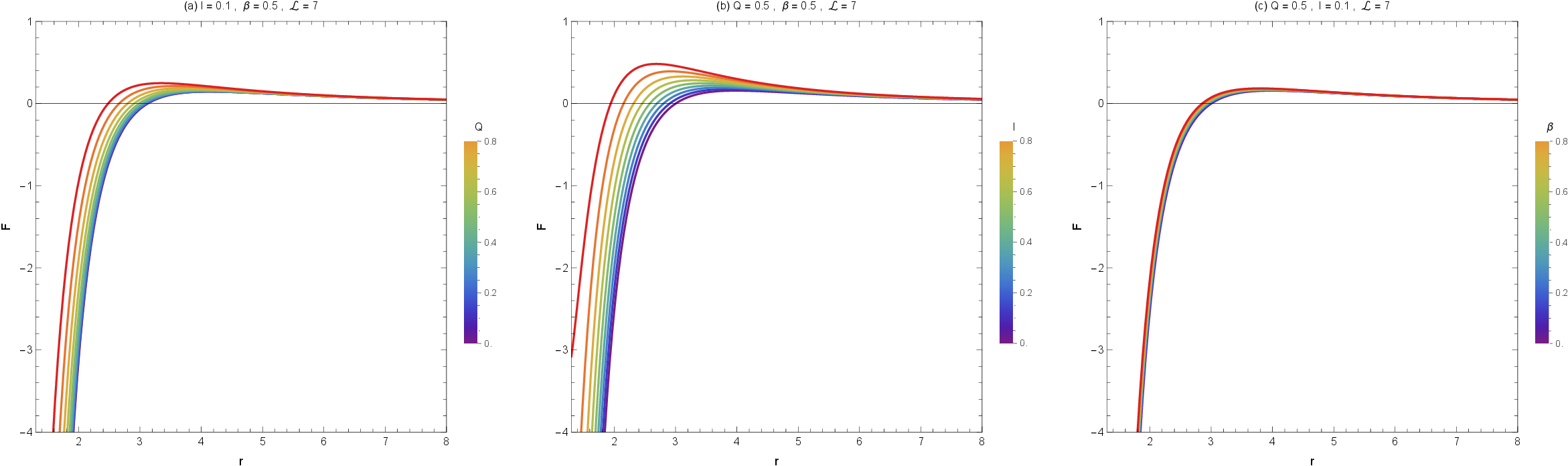}
    
    \caption{Effective force acting on particles within parameter-space ($0.0, 0.8$) of  parameters-$Q, l, \beta$.}
    \label{fig:Force}
\end{figure}
Figure~\ref{fig:Force} presents the radial dependence of the effective force $F$ for different parameter sets $(Q, l, \beta)$. It is observed that increasing the parameters $Q$ and $l$ enhances the attractive nature of the force $(F<0)$, indicating a stronger inward gravitational pull on test particles [see Fig.~\ref{fig:Force}(a, b)] whereas $\beta$ induce only marginal changes in the force profile, reflecting its comparatively weak influence on the radial dynamics [cf. fig.~\ref{fig:Force}(c)]. The force exhibits a pronounced variation in the near horizon-region, where deviations from the Schwarzschild solution behavior are most significant. At larger radial distances ($r \to \infty$), the magnitude of the force gradually decreases, approaching near-vanishing values and indicating the spacetime asymptotically approaches to a weak-field regime. Notably, the stronger attractive behavior associated with higher values of $Q$ and $l$ suggests an increased confinement of particles in the vicinity of the BH, which may affect both the stability of bound orbits and the efficiency of accretion processes. From a physical perspective, the amplification of the attractive effective force implies that the parameters $Q$ and $l$ reinforce the gravitational interaction experienced by test particles, thereby modifying orbital dynamics and potentially shifting stable orbits closer to the event horizon. This trend is consistent with the behavior observed in the effective potential and ISCO analysis, and may have further implications for accretion disk structure, epicyclic oscillations and observable signatures in strong-field astrophysical environments.
\section{Harmonic oscillations as perturbations of circular orbits in the Logarithmic KR black hole} \label{sec:HO}
Harmonic oscillations arising as small perturbations about circular geodesics provide an important  framework for analyzing the stability and dynamical response of particles in strong gravitational fields. In the vicinity of a BH, slight radial and vertical deviations from an equilibrium circular orbit lead to oscillatory motion characterized by the so-called epicyclic frequencies, whose behavior is governed by the underlying spacetime geometry~\cite{h1}. These oscillations play a crucial role in determining orbital stability and are directly linked to observable phenomena such as QPOs in accretion disks~\cite{h2}. In particular, the radial epicyclic frequency vanishes at the ISCO, marking the onset of dynamical instability. Extensions of this analysis to modified or KR BH field backgrounds, reveal measurable deviations in oscillation frequencies, thereby offering a potential avenue to test alternative gravity models through astrophysical observations~\cite{h3, h4}. Our analysis here is based on introducing small perturbations to the equations of motion in the vicinity of stable circular orbits in order to examine the resulting oscillatory behavior of test particles. When a particle is slightly displaced from its equilibrium position corresponding to a stable circular orbit in the equatorial plane, it executes epicyclic motion. In the linear approximation, this motion can be described as harmonic oscillations about the unperturbed circular trajectory.
\subsection{Local observer measurements of orbital frequencies} \label{subsec:local-freq}
The measurement of orbital frequencies by a local observer provides a physically meaningful characterization of particle dynamics and it incorporate gravitational redshift and frame-dragging effects inherent to the background field. The locally measured (proper) frequencies differ from those defined at infinity, offering a more realistic description of motion as perceived by observers comoving with the gravitational field~\cite{e2, e3}. Such local quantities are particularly important in the analysis of strong-field regions near BH, where they are directly connected to observable phenomena such as QPOs in accretion disks and these frequencies near the ISCO provides key insights into orbital stability and energy transport processes~\cite{h5, h6}.
\par
The frequencies parameters of  harmonic oscillations, as measured by a local observer, are given by
\begin{equation}
    \omega_{r}^{2}
    = -\frac{1}{2}
      \frac{\partial^{2} V_{\text{eff}}(r,\theta)}{\partial r^{2}}
    \label{eq:omega_r_def}
\end{equation}
\begin{equation}
    \omega_{\theta}^{2}
    = \frac{1}{2}
      \frac{g_{rr}}{r^{2}}
      \frac{\partial^{2} V_{\text{eff}}(r,\theta)}{\partial \theta^{2}}
    \label{eq:omega_theta_def}
\end{equation}
\begin{equation}
    \omega_{\phi} = \frac{d\phi}{d\tau}
    \label{eq:omega_phi_def}
\end{equation}
where ($\omega_{r}$, $\omega_{\theta}$, $\omega_{\phi}$)  are  the radial and angular  frequencies of  test particle. For the Logarithmic KR  BH background, these equations take the following form
\begin{align}
  \omega_{r}^{2}=\frac{1}{2} \Bigg[ 
&-\left(\frac{\mathcal{L}^2}{r^2}+1\right)
\Bigg(
\frac{6 \beta ^2 l^2 Q^2}
{r^6 \left(\frac{\beta }{r}+1\right)^2 
A^4(r)}
+ \frac{2 \beta ^2 l^2}
{r^4 \left(\frac{\beta }{r}+1\right)^2 
A^3(r)}
- \frac{2 \beta ^2 l Q^2}
{r^6 \left(\frac{\beta }{r}+1\right)^2 A^3(r)}
\nonumber \\[6pt]
&+ \frac{12 \beta l Q^2}
{r^5 \left(\frac{\beta }{r}+1\right) 
A^3(r)}
+ \frac{6 Q^2}
{r^4 A^2(r)}
- \frac{\beta ^2 l}
{r^4 \left(\frac{\beta }{r}+1\right)^2 A^2(r)}
+ \frac{2 \beta l}
{r^3 \left(\frac{\beta }{r}+1\right) 
A^2(r)}
- \frac{4}{r^3}
\Bigg)
\nonumber \\[8pt]
&- \frac{6 \mathcal{L}^2}{r^4}
\left(
\frac{Q^2}
{r^2 A^2(r)}
+ \frac{1}{A(r)}
- \frac{2}{r}
\right)
+ \frac{4 \mathcal{L}^2}{r^3}
\Bigg(
-\frac{2 \beta l Q^2}
{r^4 \left(\frac{\beta}{r}+1\right)
A^3(r)}
- \frac{2 Q^2}
{r^3 A^2(r)}
- \frac{\beta l}
{r^2 \left(\frac{\beta}{r}+1\right)
A^2(r)}
+ \frac{2}{r^2}
\Bigg)
\Bigg]
  \label{eq:omega_r2}
\end{align}
  
\begin{equation}
    \omega_{\theta}^{2}
    =
   {\frac{ -\frac{2 \beta  l Q^2}{r^4 \left(\frac{\beta }{r}+1\right) A^3(r)}-\frac{2 Q^2}{r^3 A^2(r)}-\frac{\beta  l}{r^2 \left(\frac{\beta }{r}+1\right) A^2(r)}+\frac{2}{r^2}}{ \frac{ 4 Q^2}{r A^2(r)}+\frac{2 r}{A(r)}+\frac{2 \beta  l Q^2}{r^2 \left(\frac{\beta }{r}+1\right) A^3(r)}+\frac{\beta  l}{ \left(\frac{\beta }{r}+1\right) A^2(r)}-6}} = \omega_{\phi}^{2}
    \label{eq:omega_theta2}
\end{equation}
In the limit $l=0$ and  $Q = l = \beta = 0$, Eqs.~\eqref{eq:omega_r2}–\eqref{eq:omega_theta2} reduce to the standard expressions for epicyclic frequencies of test particles in the RN and Schwarzschild BH solutions respectively~\cite{Alomar2025}.
\subsection{Frequencies measured by  distant observer} \label{subsec:distant-freq}
The locally defined orbital frequencies $\omega_i$ are given in Eqs.~\eqref{eq:omega_r2}–\eqref{eq:omega_theta2}, when these frequencies are evaluated from the perspective of a static distant observer, they are expressed in terms of the observable angular frequencies $\Omega_i$, given by
\begin{equation}
\Omega_i = \omega_i \frac{d\tau}{dt}
\end{equation}
where $ \frac{d\tau}{dt}$ represents the redshift factor, determined by
\begin{equation}
\frac{dt}{d\tau} = -\frac{\mathcal{E}}{g_{tt}}
\end{equation}
When the frequencies associated with small harmonic oscillations are measured in physical units by a distant observer, the corresponding observable frequencies of  particles can be written as
\begin{equation}
\nu_i = \frac{1}{2\pi} \frac{c^3}{GM} \, \Omega_i \; [\text{Hz}]
\end{equation}
where $i \in \{r, \theta, \phi\}$. $\Omega_r$, $\Omega_\theta$ and $\Omega_\phi$ denote respectively the dimensionless radial, latitudinal and axial angular frequencies as measured by the distant observer, given by
\begin{multline}
\Omega_{r}^{2}=
\frac{1}{2 r^6 (\beta+r)^2 
\left(l\log\!(\frac{\beta+r}{r})-1 \right)^6}
\Bigg[ Q^4\!\left(2\beta^2(3l+4)+\beta(5l+8)r+4r^2\right)+ Q^2 r \Big(2\beta^2(-6l^2+4l+9)
-3r^2(\beta l-6) 
+2\beta r((2-3\beta)l \\
+18)\Big) - l^2\left(\log\! (\frac{\beta+r}{r}) \right)^2
\Bigg[
3Q^2 r \Big(4\beta^2(l^2-2l-9)  
+r^2(\beta l-36)
+2\beta r((\beta-2)l-36)\Big) + r^2 \Big(180\beta^2 
+4\beta r\big(\beta(3l^2 \\
-18l-5)+90\big)  
 + r^3(3\beta l-20)
+2r^2(3\beta^2 l-10\beta(3l+2)+90)\Big) + 8Q^4(\beta+r)^2
\Bigg] 
+ l\log\!\left(\frac{\beta+r}{r}\right)
\Bigg[
6Q^2 r \Big(4\beta^2(l^2-l-3) 
+r^2(\beta l-12) \\
+2\beta r((\beta-1)l-12)\Big)  + r^2 \Big(72\beta^2
+2\beta r(\beta(6l^2-24l-5)+72) 
+ r^3(3\beta l-10) 
+r^2(6\beta^2 l-20\beta(2l+1)+72)\Big) +2Q^4\!\big(2\beta^2(3l+4)\\ +\beta(5l+16)r+8r^2\big)
\Bigg] 
- r^2 \Big(12\beta^2
+2\beta r(\beta(2l^2-6l-1)+12) + r^3(\beta l-2) 
+2r^2(\beta^2 l-\beta(5l+2)+6)\Big) 
+ l^3 r \left(\log\!\left(\frac{\beta+r}{r}\right)\right)^3 \\
\bigg[
r\Big(240\beta^2 +4\beta r(\beta(l^2-12l-5)+120)
+r^3(\beta l-20) 
 +2r^2(\beta^2 l-20\beta(l+1)+120)\Big) -4Q^2\!\left(2\beta^2(l+9)
+\beta(l+36)r+18r^2\right)
\bigg]
\Bigg]
 \label{eq:Omega_r2}
\end{multline}
\begin{equation}
    \Omega_{\theta}^{2}
    = \frac{Q^2 \left[\frac{\beta  l}{\beta +r}-l \log \left(\frac{\beta +r}{r}\right)+1\right]}{r^4 \left[ \log
   \left(\frac{\beta +r}{r}\right)-1\right]^3}-\frac{\beta  l}{2 r^2 (\beta +r) \left[ \log \left(\frac{\beta
   +r}{r}\right)-1\right]^2}+\frac{1}{r^3} = \Omega_{\phi}^{2}
    \label{eq:Omega_theta2}
\end{equation}
\begin{figure}[t]
    \centering
    \includegraphics[width=\linewidth]{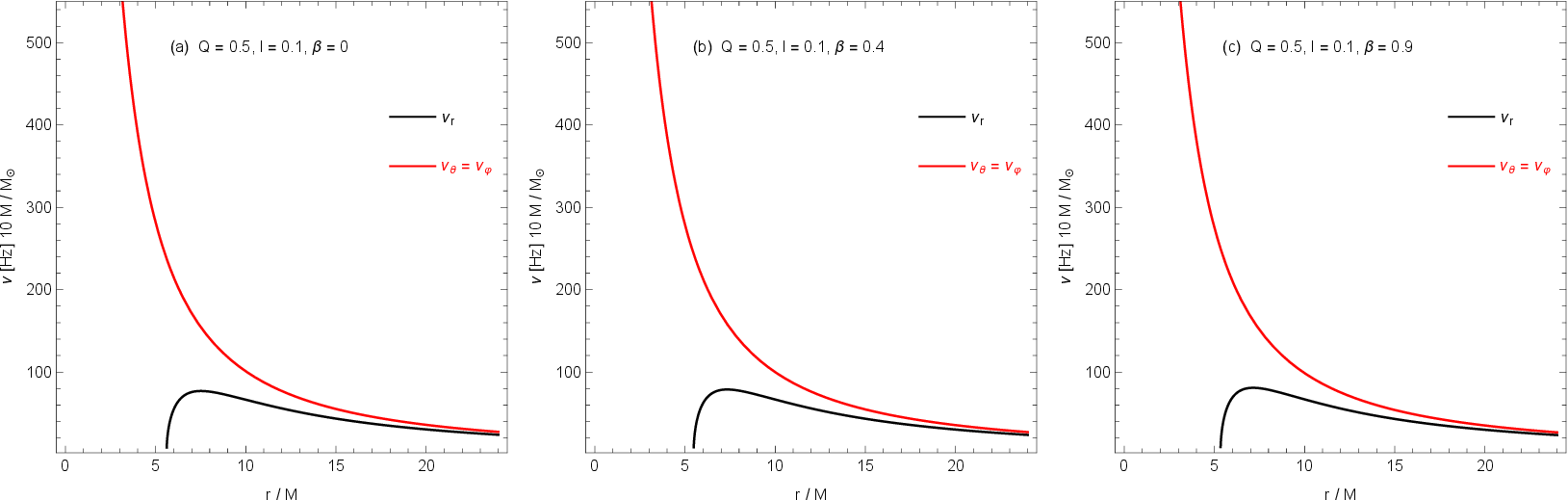}\\
    \vspace{0.3cm}
    
    \includegraphics[width=\linewidth]{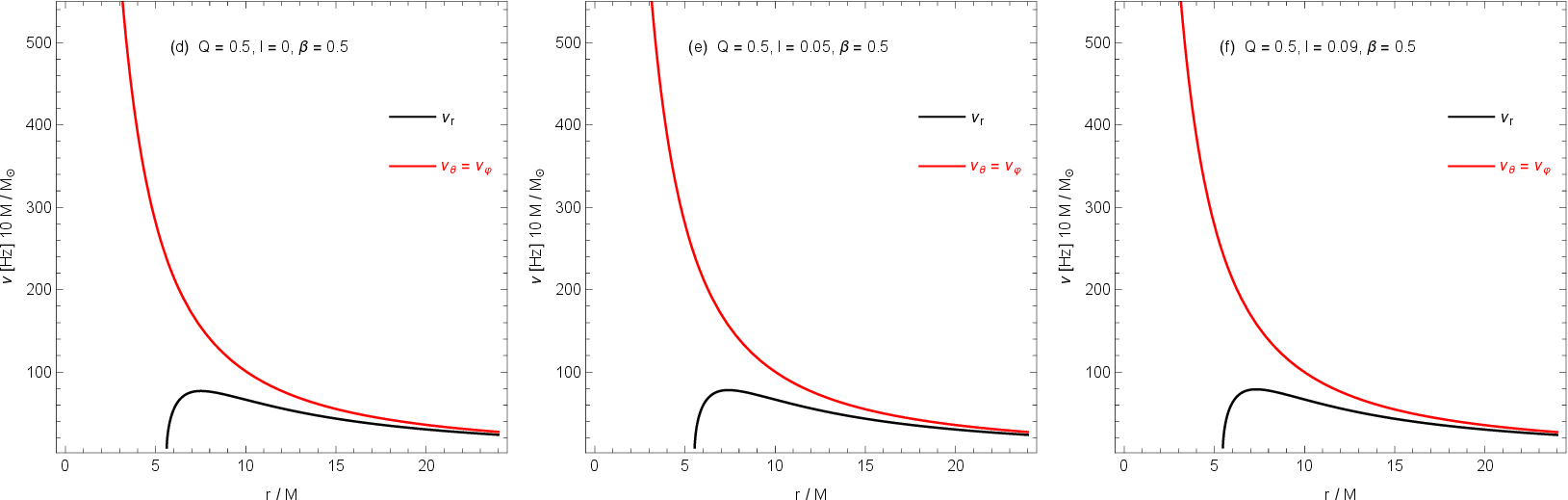}\\
    \vspace{0.3cm}
    
    \includegraphics[width=\linewidth]{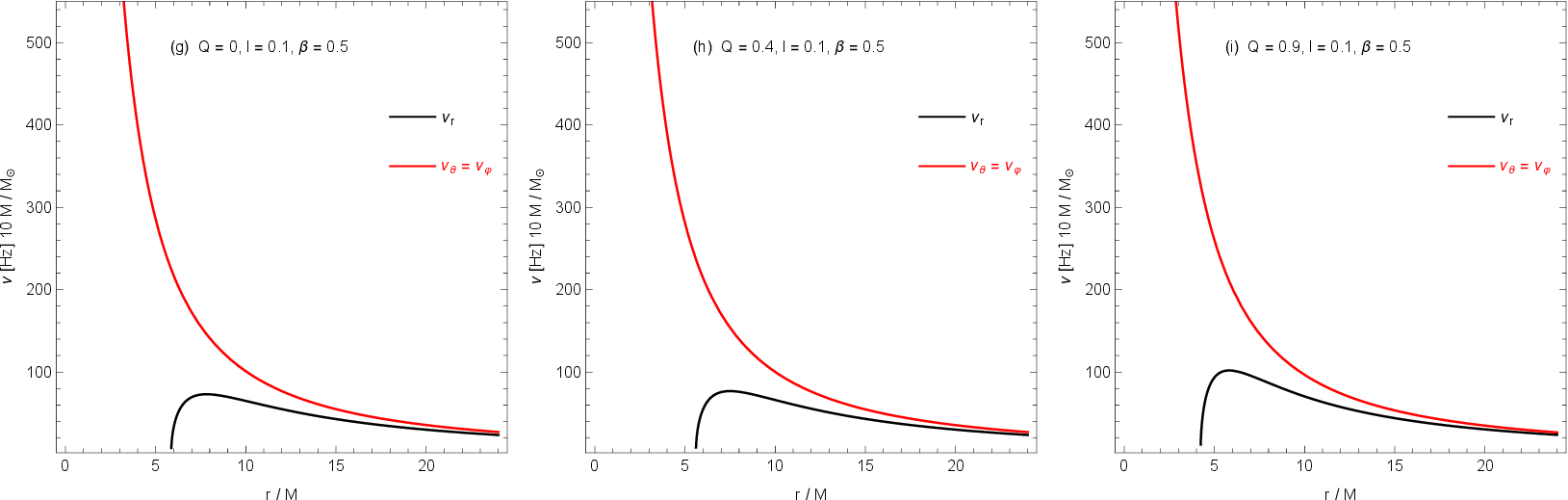}\\
    \caption{Profiles of the oscillation frequencies $(\nu_{r}, \nu_{\theta}, \nu_{\phi})$ of  particles around Logarithmic KR BH within parameter-space ($0.0, 0.8$) of  parameters-$Q, l, \beta$ measured
    by a distant observer.}
    \label{fig:freq_profiles_a_bc}
\end{figure}
Figures~\ref{fig:freq_profiles_a_bc} present the radial profiles of the oscillation frequencies $(\nu_r, \nu_\theta, \nu_\phi)$ corresponding to small harmonic perturbations of particles orbiting the Logarithmic KR BH, as measured by a distant observer. The dependence of these frequencies on the parameters $Q, l$ and $\beta$ is systematically displayed. From  \textit{ panels (a–f)}, it is evident that variations in $l$ and $\beta$ induce small modifications in the frequency behavior, whereas \textit{ panels (g–i)} show that increasing the charge parameter $Q$ leads to a pronounced inward shift of the frequency profiles toward the event horizon. These results indicate that the characteristic QPOs frequencies are sensitive to the underlying parameters, with their combined influence producing noticeable deviations from standard profiles. The radial epicyclic frequency $\nu_r$ exhibits a non-monotonic behavior- it vanishes near the ISCO, attains a maximum at intermediate radii, and subsequently decreases at larger distances. The enhancement of its peak amplitude with increasing $Q$ suggests that the BH charge exerts the strongest influence on radial oscillatory motion. This behavior reflects the stability properties of radial perturbations and highlights their sensitivity to the presence of the KR field. The vertical and azimuthal frequencies, $\nu_\theta$ and $\nu_\phi$, display nearly identical behavior across the parameter space, decreasing monotonically with radial distance and remaining effectively degenerate. In the absence of rotation, there is no frame-dragging mechanism to lift this degeneracy, leading to $\nu_\theta \approx \nu_\phi$. From a physical perspective, the distinct behaviors of $\nu_r$ and $\nu_\theta$ are particularly relevant for QPO models, as differences among these frequencies govern relativistic precession and resonance phenomena. The comparatively stronger sensitivity of $\nu_r$ to the BH parameters suggests that radial oscillations may provide the most robust observational signature of the logarithmic KR BH.
 
\subsection{Periastron precession} \label{subsec:peri}
Periastron precession is a fundamental relativistic effect describing the advance of the point of closest approach in bound orbits and  arises due to gravitational fields, leading to a mismatch between the axial angular frequencies and radial frequencies of orbital motion~\cite{e3}. For test particles orbiting BHs, the precession rate depends sensitively on the underlying metric parameters and becomes particularly pronounced in the strong-field regime near the ISCO. This also plays a key role in the interpretation of QPOs and orbital dynamics in accretion disks around BH~\cite{ p1}.
\par
In the presence of the KR field, deviations in the periastron advance may serve as significant observational indicators for distinguishing such configurations from standard BH solutions. Accordingly, we investigate the periastron precession frequency of a  test particle orbiting a logarithmic KR BH, with particular focus on small perturbations about the equatorial plane $\theta=\pi/2$. To this end, a slight displacement from the equilibrium circular orbit is introduced, leading to oscillatory motion governed by the radial frequency $\Omega_r$. The periastron precession frequency $\Omega_P$, is then defined as the difference between the axial angular frequencies and the radial frequencies,

\begin{equation}
    \Omega_{P} = \Omega_{\phi} - \Omega_{r}.
    \label{eq:OmegaP_def}
\end{equation}
where $\Omega_r$ and $\Omega_\phi$ are given in Eqs. (\ref{eq:Omega_r2})-(\ref{eq:Omega_theta2}).
\begin{figure}[t]
    \centering
    \includegraphics[width=\linewidth]{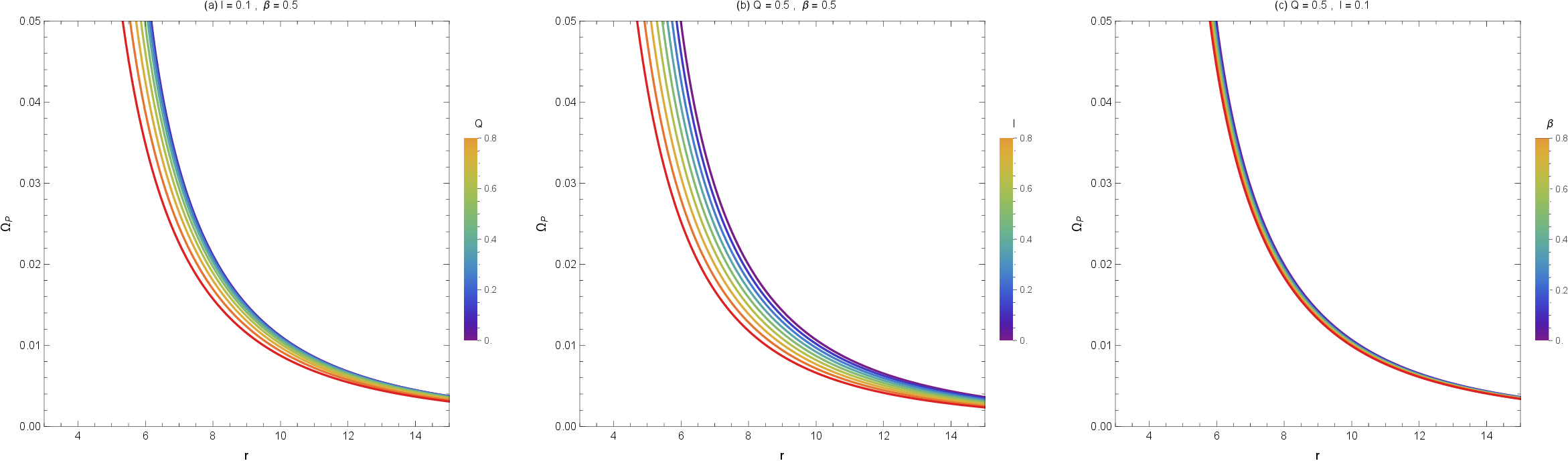}
    \caption{Periastron precession frequency $\Omega_{P}$ of  test particles
    orbiting a Logarithmic KR BH  within parameter-space ($0.0, 0.8$) of  parameters-$Q, l, \beta$.}
    \label{fig:OmegaP}
\end{figure}
Figure~\ref{fig:OmegaP} displays the radial dependence of the periastron precession frequency for test particles orbiting the logarithmic KR BH, with explicit emphasis on the roles of the parameters $Q, l$ and $\beta$. The profiles demonstrate a monotonic decrease of the precession frequency with increasing radial coordinate $r$ for all parameter configurations. Furthermore, increasing $Q$ and $l$ leads to an additional suppression of the precession frequency, underscoring their pronounced influence on the orbital dynamics. The variations in $\beta$ induce only small modifications, indicating that its contribution to the overall behavior of the Periastron precession frequency is comparatively weak.

\section{Energy Emission and Hawking Radiation} \label{sec:EEHR}
Energy emission from BHs, most notably through Hawking radiation, represents a profound connection between quantum theory, thermodynamics and gravitation. In the framework of quantum field theory in curved Spacetime, particle creation arises due to quantum fluctuations near the event horizon, leading to a thermal radiation spectrum characterized by the BH temperature. This emission results in a gradual loss of mass and energy, implying that BHs are not entirely black but evolve over time through evaporation.  As a result, the BH gradually loses mass over time, eventually leading to its complete evaporation within a finite duration. The evaporation rate is directly proportional to the rate at which energy is emitted and for the distant observer, the high-energy absorption cross section closely resembles the apparent shadow cast by the BH~\cite{em3, em4}. The properties of Hawking radiation depend sensitively on the underlying fields, can alter the emission spectrum and evaporation rate. Quantum fluctuations in the vicinity of a BH give rise to the continuous creation and annihilation of particle pairs just outside the event horizon~\cite{em1, em2}. Through the quantum tunneling process, one particle-particularly the positively charged component-can be drawn toward the BH’s center, while its counterpart escapes to infinity. Consequently, studying energy emission in alternative BH models provides a potential avenue to test deviations from classical GR and to explore signatures of quantum gravity effects~\cite{108, 113}.
\par
The absorption cross section exhibits oscillatory behavior around a constant limiting value, $\sigma_{\text{lim}}$, which corresponds to the characteristic radius of the BH \cite{114},

\begin{equation}
\sigma_{\text{lim}} \approx \pi r_0^{2},
\end{equation}
where $r_0$ denotes the radius of the event horizon, determined from the condition $f(r_0)=0$.
\par
Thus, the expression for the energy emission rate of the BH is given by
\begin{equation}
\frac{d^{2}\varepsilon}{d\omega dt}
=
\frac{2\pi^{2} \sigma_{\text{lim}} \, \omega^{3}}
{e^{\omega/T_{H}} - 1},
\label{Energy E}
\end{equation}
where $\omega$ represents the radiation frequency and $T_{H}$ is the Hawking temperature of the BH. The Hawking temperature is determined by the surface gravity at the event horizon and can be expressed as
\begin{equation}
T_{H} = \frac{1}{4\pi}
\left.
\frac{\partial f(r)}{\partial r}
\right|_{r=r_0}
\label{Th}
\end{equation}
After substituting the explicit expression of $f(r)$ into the above relation and evaluating the derivative at $r=r_0$, the Hawking temperature (\ref{Th}) reduces to the form
\begin{equation}
T_{H} =
\frac{1}{4\pi}
\left[
{-\frac{2 \beta  l Q^2}{r_{0}^4 \left(\frac{\beta }{r_0}+1\right) A_0^3}-\frac{2 Q^2}{r_0^3 A^2_0}-\frac{\beta 
   l}{r_0^2 \left(\frac{\beta }{r_0}+1\right) A^2_0}+\frac{2}{r_0^2}}
\right].
\end{equation}
where $A_0= A(r_0)$. Substituting the expressions for the cross section  $\sigma_{\text{lim}}$ and $T_H$ into Eq.~\eqref{Energy E}, we obtain the expression for the energy emission rate in Logarithmic KR BH,

\begin{equation}
\begin{aligned}
\frac{d^{2}\varepsilon}{d\omega dt}
&=
\left( 2\pi^{3} r_{0}^{2} \right)\omega^{3}
\Bigg[
\exp\Bigg(
\omega
\Bigg\{
\frac{1}{4\pi}
\Bigg[
-\frac{2 \beta l Q^2}{r_0^4 \left(\frac{\beta}{r_0}+1\right) A^3_0}
-\frac{2 Q^2}{r_0^3 A^2_0}
\\
&\qquad
-\frac{\beta l}{r_0^2 \left(\frac{\beta}{r_0}+1\right) A^2_0}
+\frac{2}{r_0^2}
\Bigg]
\Bigg\}^{-1}
\Bigg)
- 1
\Bigg]^{-1} 
\end{aligned}
\end{equation}
The energy emission rate, derived from the associated Hawking radiation temperature, encodes the effects of quantum and thermal fluctuations on BH radiative processes. The inclusion of logarithmic corrections reduces the Hawking temperature, thereby suppressing the emission rate and indicating a slower evaporation timescale. This behavior points to an enhancement of thermodynamic stability, particularly in the small-mass regime.
\par
\begin{figure}[t]
    \centering
    \includegraphics[width=\linewidth]{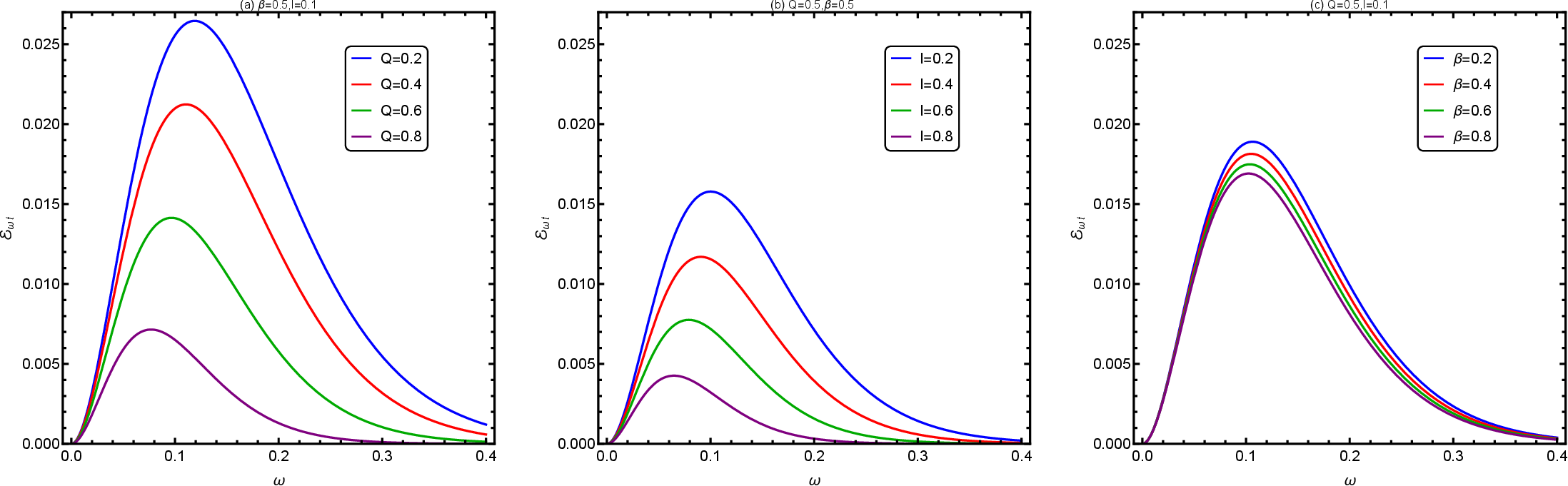}

    \caption{Energy emission rate $\varepsilon_{\omega t}$ in Logarithmic KR BH with variation of parameters $Q$,$l$ and $\beta$}
    \label{fig:EE}
\end{figure}

A detailed analysis of these features is presented in Fig.~\ref{fig:EE} (a–c), where the spectral energy emission rate $\varepsilon_{\omega t} (= \frac{d^{2}\varepsilon}{d\omega dt})$ is shown to decrease systematically with increasing values of the parameters $Q, l$ and $\beta$, demonstrating their direct influence on the radiative properties of the system. Furthermore, the emission spectra exhibit a pronounced peak at intermediate frequencies $(\omega)$, followed by a gradual decline at higher frequencies, indicating the presence of a characteristic frequency band where the radiation output is maximized. The observed leftward shift and variation in peak amplitude suggest that these parameters affect both the intensity and spectral distribution of the emitted radiation. In particular, larger values of $Q$ and $l$ tend to broaden the emission profile and modify its amplitude, reflecting enhanced thermal activity associated with these parameter regimes. From a physical perspective, these results imply that the parameters $Q, l$  and $\beta$ significantly influence the thermodynamic behavior of the Logarithmic KR BH through modifications of its radiation spectrum. The shifts in the emission peak can be attributed to changes in the effective temperature and the underlying quantum-corrected geometry. Such features may have important implications for BH evaporation, stability, and potential observational signatures, especially in the context of primordial or modified gravity BH, where suppressed emission rates could play a key role in determining their detectability in high-energy astrophysical environments.

\section{Conclusions and Summary of the work} \label{sec:conclusion}
We have introduced a new class of BH solutions by introducing a logarithmic deformation parameter of the KR field, thereby formulating the Logarithmic KR BH model. This work extends conventional linear and algebraic approaches by incorporating two independent parameters- the Lorentz-violating parameter $l$( which encodes the deviations from local Lorentz symmetry) and the logarithmic coupling parameter $\beta$, which captures purely nonlinear interactions of the KR field. The horizon analysis further reveals a nontrivial parameter-dependent structure of the logarithmic KR BH. Within the admissible parameter space, the metric function $f(r)$ admits two distinct positive roots corresponding to  $r_{-}$ (inner Cauchy horizon) and $r_{+}$ (outer event horizon). The larger values of BH parameters $Q$ and  $l$ eliminate the horizon solutions whereas the  parameter $\beta$ plays an important role in determining the location and existence of the horizons, thereby providing an additional degree of freedom for controlling the causal structure of the BH spacetime. We have analyzed the physical and dynamical properties of our BH model, demonstrating that the combined influence of $Q, l$ and $\beta $ leads to significant modifications in the horizon structure and the effective potential governing particle motion. These parameters  play a dominant role in deforming the spacetime geometry, shifting the horizon locations and altering the causal structure. Our proposed model modifies directly impact the stability of circular orbits and the location of the ISCO, indicating that stable bound orbits can exist closer to the BH compared to standard solutions. The dynamics of test particles have been investigated using the effective potential formalism, allowing us to derive analytical expressions for the specific energy and angular momentum and show that the parameters $Q$ and $l$ significantly influence orbital stability and energetics. Furthermore, the study of epicyclic motion reveals that the radial, latitudnal and axial angular frequencies are sensitive to the underlying filed structure, with observable implications for QPOs. Although the present work does not directly compare the obtained results with observational QPO data, it establishes a theoretical framework that may help distinguish this BH model from classical ones through their QPO behavior. Moreover, advancements in high-resolution X-ray observations could enable detailed analyses of these QPO features, offering further opportunities to test the predictions of the model. The analysis of Periastron precession further demonstrates that deviations from standard BH models can be substantial, particularly in the strong-field region, thereby providing potential probes of modified gravity effects.
\par
From a thermodynamic perspective, we have examined the impact of logarithmic corrections on the Hawking radiation temperature and the associated energy emission rate. The presence of the deformation parameter$(\beta)$ leads to a suppression of the Hawking temperature and consequently reduces the emission rate, implying a slower evaporation process and enhanced thermodynamic stability, especially for small-mass BHs. The spectral analysis of the emission rate further indicates that the parameters  $Q, l$ and $\beta $ modify both the intensity and distribution of radiation, with possible implications for observational signatures in high-energy astrophysical environments. Overall, the present study demonstrates that Logarithmic KR BH exhibit clear departures from classical solutions such as the Schwarzschild solution and Reissner-Nordstrom solution. The interplay between Lorentz-violating effects and nonlinear KR couplings introduces rich phenomenology, affecting horizon structure, geodesic stability, oscillatory dynamics, and thermodynamic behavior. The results provide important insights into the complex relationship between BH properties and particle dynamics, offering potential approaches for testing these theoretical models through astrophysical observations. Our findings contribute to a deeper understanding of the Logarithmic KR BH and its influence on the surrounding spacetime environment. These findings also suggest that our model provide a viable and physically motivated extension of BH physics, with potential relevance for testing deviations from GR through astrophysical observations and  by extending this analysis to rotating Logarithmic KR BHs could further clarify the role of spin and additional parameters in shaping particle dynamics and observable effects.
\par
\textbf{Funding Information:} No funding was received for this research.
\par
\textbf{Data Availability Statement:}  No or not applicable. This manuscript does not report data generation or analysis.
\par
\textbf{Declaration of Competing Interest:} The authors confirm that there are no financial interests or personal affiliations that could have influenced the research presented in this paper.
\par
\textbf{Author Contribution: } Aftab Ansari(AA): Conceptualization, Formal analysis, Methodology, Writing-original draft. Rajesh Kumar(RK): Conceptualization, Methodology, Investigation, Validation, Review, Editing, Supervision and Project administration. Praveen Kumar Dhankar: Supervision, Conceptualization, Validation, Review and Editing.
\par
\textbf{Acknowledgment:} The author AA acknowledge the CSIR, Government of India, for support via Junior Research Fellowship with File No. 09/0057(25433)/2025-EMR-I and ID No. 35113857. The authors RK is thankful to IUCAA, Pune, India, for providing facilities under associateship programs where a part of the work is done during their visit.


\begin{thebibliography}{99}

\bibitem{1} Einstein, A. (1911). On the Influence of Gravitation on the Propagation of Light. Annalen der Physik, 35(10), 898-908.
\bibitem{2}Will, C. M. (2014). The confrontation between general relativity and experiment. Living reviews in relativity, 17(1), 1-117.
\bibitem{3} Schwarzschild, K. (1916). Über das gravitationsfeld eines massenpunktes nach der einsteinschen theorie. Sitzungsberichte der königlich preussischen Akademie der Wissenschaften, 189-196.



\bibitem{9}Falcke, H., Melia, F., \& Agol, E. (2000). Viewing the shadow of the black hole at the galactic center. The Astrophysical Journal Letters, 528(1), L13-L16.
\bibitem{10}Synge, J. L. (1966). The escape of photons from gravitationally intense stars. Monthly Notices of the Royal Astronomical Society, 131(3), 463-466.
\bibitem{11}Event Horizon Telescope Collaboration, Akiyama, K., Alberdi, A., Alef, W., Asada, K., Azulay, R., ... \& Pu, H. Y. (2019). First M87 event horizon telescope results. II. Array and instrumentation. The Astrophysical Journal Letters, 875(1), L2.
\bibitem{12}Akiyama, K., Alberdi, A., Alef, W., Asada, K., Azulay, R., Baczko, A. K., ... \& Ramakrishnan, V. (2019). First M87 event horizon telescope results. III. Data processing and calibration. The Astrophysical Journal Letters, 875(1), L3.
\bibitem{13}Event Horizon Telescope Collaboration, Akiyama, K., Alberdi, A., Alef, W., Asada, K., Azulay, R., ... \& Ramakrishnan, V. (2019). First M87 event horizon telescope results. V. Physical origin of the asymmetric ring. The Astrophysical Journal Letters, 875(1), L5.
\bibitem{14}Event Horizon Telescope Collaboration, Akiyama, K., Alberdi, A., Alef, W., Asada, K., Azulay, R., ... \& Ramakrishnan, V. (2019). First M87 event horizon telescope results. VI. The shadow and mass of the central black hole. The Astrophysical Journal Letters, 875(1), L6.

\bibitem{15}Akiyama, K., Alberdi, A., Alef, W., Algaba, J. C., Anantua, R., Asada, K., ... \& Marti-Vidal, I. (2022). First Sagittarius A* Event Horizon Telescope results. I. The shadow of the supermassive black hole in the center of the Milky Way. The Astrophysical Journal Letters, 930(2), L12.
\bibitem{16}Event Horizon Telescope Collaboration, Akiyama, K., Alberdi, A., Alef, W., Algaba, J. C., Anantua, R., ... \& Marscher, A. P. (2022). First Sagittarius A* Event Horizon Telescope results. IV. Variability, morphology, and black hole mass. The Astrophysical Journal Letters, 930(2), L15.
\bibitem{17}Event Horizon Telescope Collaboration, Akiyama, K., Alberdi, A., Alef, W., Algaba, J. C., Anantua, R., ... \& Marscher, A. P. (2022). First Sagittarius A* event horizon telescope results. VI. Testing the black hole metric. The Astrophysical Journal Letters, 930(2), L17.
\bibitem{18}Cardoso, V., \& Pani, P. (2019). Testing the nature of dark compact objects: a status report. Living Reviews in Relativity, 22(1), 4.
\bibitem{19}Zahid, M., Khan, S. U., Ren, J., \& Rayimbaev, J. (2022). Geodesics and shadow formed by a rotating Gauss–Bonnet black hole in AdS spacetime. International Journal of Modern Physics D, 31(08), 2250058.

\bibitem{21} Zahid, M., Sarikulov, F., Shen, C., Umaraliyev, M., \& Rayimbaev, J. (2024). Shadow and quasinormal modes of novel charged rotating black hole in Born–Infeld theory: constraints from EHT results. Physics of the Dark Universe, 46, 101616.
\bibitem{22}Khan, S. U., \& Ren, J. (2020). Shadow cast by a rotating charged black hole in quintessential dark energy. Physics of the Dark Universe, 30, 100644.
\bibitem{23}Khan, S. U., \& Ren, J. (2022). Geodesics and optical properties of a rotating black hole in Randall–Sundrum brane with a cosmological constant. Chinese Journal of Physics, 78, 141-154.

\bibitem{24}Zubair, M., Raza, M. A., \& Maqsood, E. Physics of the Dark Universe.
\bibitem{25}Raza, M. A., Rayimbaev, J., Sarikulov, F., Zubair, M., Ahmedov, B., \& Stuchlík, Z. (2024). Shadow of novel rotating black hole in GR coupled to nonlinear electrodynamics and constraints from EHT results. Physics of the Dark Universe, 44, 101488.
\bibitem{26}Rayimbaev, J., Tadjimuratov, P., Abdujabbarov, A., Ahmedov, B., \& Khudoyberdieva, M. (2020). Dynamics of test particles around regular black holes in modified gravity. arXiv preprint arXiv:2010.12863.
\bibitem{27}Ahmed, F., Al-Badawi, A., \& Sakallı, İ. Quantum Oppenheimer–Snyder Ads Black Hole with Quintessential Dark Energy and String Clouds Geodesics, Perturbative Dynamics, and Thermal Properties. Perturbative Dynamics, and Thermal Properties.
\bibitem{28}Rink, K., Caiazzo, I., \& Heyl, J. (2022). Testing general relativity using quasi-periodic oscillations from X-ray black holes: XTE J1550-564 and GRO J1655-40. Monthly Notices of the Royal Astronomical Society, 517(1), 1389-1397.

\bibitem{29}Pourhassan, B., Dehghani, M., Upadhyay, S., Sakallı, İ., \& Singh, D. V. (2022). Exponential corrected thermodynamics of Born–Infeld BTZ black holes in massive gravity. Modern Physics Letters A, 37(33n34), 2250230.
\bibitem{30}Mustafa, G., Hussain, I., \& Liu, W. M. (2021). Quasi periodic oscillations of test particles and red-blue shifts of the photons emitted by the charged test particles orbiting the charged black hole in the presence of quintessence and clouds of strings. arXiv preprint arXiv:2108.07801.
\bibitem{31}Sakalli, I., \& Kanzi, S. (2022). Topical review: greybody factors and quasinormal modes for black holes invarious theories-fingerprints of invisibles. Turkish Journal of Physics, 46(2), 51-103.
\bibitem{32}Huang, G. Y., \& Nath, N. (2022). Neutrino meets ultralight dark matter:decay and cosmology. Journal of Cosmology and Astroparticle Physics, 2022(05), 034.
\bibitem{33}Ahmed, F., Al-Badawi, A., Sakallı, İ., \& Shaymatov, S. (2025). Dynamics of test particles and scalar perturbation around an Ayón–Beato–García black hole coupled with a cloud of strings. Chinese Journal of Physics, 96, 770-791.
\bibitem{81}Homan, J., Klein-Wolt, M., Rossi, S., Miller, J. M., Wijnands, R., Belloni, T., ... \& Lewin, W. H. (2003). High-frequency quasi-periodic oscillations in the black hole x-ray transient xte j1650–500. The Astrophysical Journal, 586(2), 1262-1267.

\bibitem{84}Rezzolla, L., \& Zanotti, O. (2013). Relativistic hydrodynamics. Oxford University Press.

\bibitem{HFQPOs and LFQPOs1}Belloni, T. M., Bhattacharya, D., Caccese, P., Bhalerao, V., Vadawale, S., \& Yadav, J. S. (2019). A variable-frequency HFQPO in GRS 1915+ 105 as observed with AstroSat. Monthly Notices of the Royal Astronomical Society, 489(1), 1037-1043.
\bibitem{HFQPOs and LFQPOs2}Remillard, R. A. (2004, July). X‐ray QPOs from Black Hole Binary Systems. In AIP Conference Proceedings (Vol. 714, No. 1, pp. 13-20). American Institute of Physics.
\bibitem{85}Rezzolla, L., Yoshida, S. I., \& Zanotti, O. (2003). Oscillations of vertically integrated relativistic tori–I. Axisymmetric modes in a Schwarzschild space-time. Monthly Notices of the Royal Astronomical Society, 344(3), 978-992.
\bibitem{86}Amarilla, L., Eiroa, E. F., \& Giribet, G. (2010). Null geodesics and shadow of a rotating black hole in extended Chern-Simons modified gravity. Physical Review D—Particles, Fields, Gravitation, and Cosmology, 81(12), 124045.
\bibitem{87}Stuchlík, Z., Kotrlova, A., \& Török, G. (2011). Resonant radii of kHz quasi-periodic oscillations in Keplerian discs orbiting neutron stars. Astronomy \& Astrophysics, 525, A82.
\bibitem{88}Török, G., Kotrlová, A., Šrámková, E., \& Stuchlik, Z. (2011). Confronting the models of 3: 2 quasiperiodic oscillations with the rapid spin of the microquasar GRS 1915+ 105. Astronomy \& Astrophysics, 531, A59.
\bibitem{89}Stuchlík, Z., Kotrlová, A., \& Török, G. (2013). Multi-resonance orbital model of high-frequency quasi-periodic oscillations: possible high-precision determination of black hole and neutron star spin. Astronomy \& Astrophysics, 552, A10.
\bibitem{90}Jiang, X., Wang, P., Wu, H., \& Yang, H. (2021). Testing Kerr black hole mimickers with quasi-periodic oscillations from GRO J1655-40. The European Physical Journal C, 81(11), 1043.
\bibitem{91}Mustafa, G., Hussain, I., \& Liu, W. M. (2022). Quasi-periodic oscillations of test particles and red–blue shifts of photons in the charged-Kiselev black hole with cloud of strings. Chinese Journal of Physics, 80, 148-166.
\bibitem{92}Liu, Y., Mustafa, G., Maurya, S. K., \& Javed, F. (2023). Orbital motion and quasi-periodic oscillations with periastron and Lense–Thirring precession of slowly rotating Einstein–Æther black hole. The European Physical Journal C, 83(7), 584.
\bibitem{82}Rayimbaev, J., Ahmedov, B., \& Bokhari, A. H. (2022). Constraints on charged black hole parameters using quasiperiodic oscillations data. International Journal of Modern Physics D, 31(11), 2240004.
\bibitem{83}Qi, M., Rayimbaev, J., \& Ahmedov, B. (2023). Charged particles and quasiperiodic oscillations around magnetized Schwarzschild black holes. The European Physical Journal C, 83(8), 730.
\bibitem{93}Ingram, A. R., \& Motta, S. E. (2019). A review of quasi-periodic oscillations from black hole X-ray binaries: observation and theory. New Astronomy Reviews, 85, 101524.

\bibitem{94}Kološ, M., Shahzadi, M., \& Stuchlík, Z. (2020). Quasi-periodic oscillations around Kerr-MOG black holes. The European Physical Journal C, 80(2), 133.
\bibitem{95}Shahzadi, M., Kološ, M., Stuchlík, Z., \& Habib, Y. (2021). Epicyclic oscillations in spinning particle motion around Kerr black hole applied in models fitting the quasi-periodic oscillations observed in microquasars and AGNs. The European Physical Journal C, 81(12), 1067.
\bibitem{96}Shahzadi, M., Kološ, M., Stuchlík, Z., \& Habib, Y. (2022). Testing alternative theories of gravity by fitting the hot-spot data of Sgr A. The European Physical Journal C, 82(5), 407.
\bibitem{97}Kološ, M., Shahzadi, M., \& Tursunov, A. (2023). Charged particle dynamics in parabolic magnetosphere around Schwarzschild black hole. The European Physical Journal C, 83(4), 323.
\bibitem{98}Shahzadi, M., Kološ, M., Saleem, R., Habib, Y., \& Eduarte-Rojas, A. (2023). Structure-preserving numerical simulations of test particle dynamics around slowly rotating neutron stars within the Hartle-Thorne approach. Physical Review D, 108(10), 103006.
\bibitem{99}Naseer, T. (2024). Complexity and isotropization based extended models in the context of electromagnetic field: an implication of minimal gravitational decoupling. The European Physical Journal C, 84(12), 1256.
\bibitem{100}Çalışkan, M., Chen, Y., Dai, L., Kumar, N. A., Stomberg, I., \& Xue, X. (2024). Dissecting the stochastic gravitational wave background with astrometry. Journal of Cosmology and Astroparticle Physics, 2024(05), 030.
\bibitem{101}Rasheed, B., Ditta, A., Naseer, T., Javed, F., \& Mustafa, G. (2025). Analyzing the quantum corrected adS spherically symmetric black holes with phantom global monopoles for thermal properties. International Journal of Geometric Methods in Modern Physics, 22(04), 2450302.
\bibitem{102}Ashraf, A., Ditta, A., Naseer, T., Maurya, S. K., Ray, S., Channuie, P., \& Atamurotov, F. (2025). Testing of QPOs, particle dynamics, emission energy and thermal fluctuation around a regular hairy black hole. The European Physical Journal C, 85(6), 633.
\bibitem{103}Liu, Y., Mustafa, G., Maurya, S. K., \& Javed, F. (2023). Orbital motion and quasi-periodic oscillations with periastron and Lense–Thirring precession of slowly rotating Einstein–Æther black hole. The European Physical Journal C, 83(7), 584.

\bibitem{intro1} Yang, K., Chen, Y. Z., Duan, Z. Q., \& Zhao, J. Y. (2023). Static and spherically symmetric black holes in gravity with a background Kalb-Ramond field. Physical Review D, 108(12), 124004.


\bibitem{Mukhopadhyaya2004} Mukhopadhyaya, B., Sen, S., SenGupta, S., \& Sur, S. (2004). Parity violation and torsion: A study in four and higher dimensions. arXiv preprint hep-th/0207165.
\bibitem{Kostelecky2009} Kostelecký, V. A., \& Tasson, J. D. (2009). Prospects for large relativity violations in matter-gravity couplings. Physical review letters, 102(1), 010402.
\bibitem{SenGupta2001} SenGupta, S., \& Sur, S. (2001). Spherically symmetric solutions of gravitational field equations in Kalb–Ramond background. Physics Letters B, 521(3-4), 350-356.
\bibitem{RS2025} V. P, R., \& R, S. (2025). Strong Gravitational Lensing and Shadow of the Kalb-Ramond Black Hole in Homogeneous and Inhomogeneous Plasma. Brazilian Journal of Physics, 55(6), 290.
\bibitem{Paul2020}Paul, T. (2020). Antisymmetric tensor fields in modified gravity: A summary. Symmetry, 12(9), 1573.
\bibitem{intro2} Duan, Z. Q., Zhao, J. Y., \& Yang, K. (2024). Electrically charged black holes in gravity with a background Kalb–Ramond field. The European Physical Journal C, 84(8), 798.
\bibitem{intro3} Pantig, R. C., \& Ovgun, A. (2025). Multimodal signatures of asymptotic (A) dS Kalb–Ramond black holes: Constraints through the shadow, weak deflection angle, and topological photon spheres. Annals of Physics, 480, 170104.
\bibitem{Shapiro2002} Shapiro, I. L. (2002). Physical aspects of the space–time torsion. Physics Reports, 357(2), 113-213..
\bibitem{Hammond2002}Hammond, R. T. (2002). Torsion gravity. Reports on Progress in Physics, 65(5), 599-649.
\bibitem{Paul2019} Paul, T., \& SenGupta, S. (2019). Dynamical suppression of spacetime torsion. The European Physical Journal C, 79(7), 591.

\bibitem{Obukhov2014} Puetzfeld, D., \& Obukhov, Y. N. (2014). Prospects of detecting spacetime torsion. International Journal of Modern Physics D, 23(12), 1442004.
\bibitem{intro4} Sucu, E., \& Sakallı, I. (2025). Exploring Lorentz-violating effects of Kalb-Ramond field on charged black hole thermodynamics and photon dynamics. Physical Review D, 111(6), 064049.
\bibitem{intro5} Mangut, M., Gursel, H., \& Sakallı, İ. (2025). Lorentz-symmetry violation in charged black-hole thermodynamics and gravitational lensing: effects of the Kalb-Ramond field. Chinese Physics C, 49(6), 065106.
\bibitem{intro6} Asrat, M. (2025). Kalb-Ramond field, black holes and black strings in (2+ 1) D. Journal of High Energy Physics, 2025(8), 1-53.
\bibitem{intro7} Jumaniyozov, S., Murodov, S., Rayimbaev, J., Ibragimov, I., Madaminov, B., Urinbaev, S., \& Abdujabbarov, A. (2025). Black holes surrounded by PFDM in Kalb-Ramond gravity: from thermodynamics to QPO tests. The European Physical Journal C, 85(7), 797.


\bibitem{Ingram2019} Ingram, A. R., \& Motta, S. E. (2019). A review of quasi-periodic oscillations from black hole X-ray binaries: observation and theory. New Astronomy Reviews, 85, 101524.
\bibitem{Remillard2006} Remillard, R. A., \& McClintock, J. E. (2006). X-ray properties of black-hole binaries. Annu. Rev. Astron. Astrophys., 44(1), 49-92.
\bibitem{Kato1998} Kato, S., Fukue, J., \& Mineshige, S. (1998). Black-hole accretion disks.
\bibitem{Stella1999} Stella, L., \& Vietri, M. (1999). kHz quasiperiodic oscillations in low-mass X-ray binaries as probes of general relativity in the strong-field regime. Physical Review Letters, 82(1), 17.
\bibitem{Bambi2017} Bambi, C. (2017). Black holes: a laboratory for testing strong gravity (Vol. 10, pp. 978-981). Singapore: Springer.
\bibitem{Shafee2006}Shafee, R., McClintock, J. E., Narayan, R., Davis, S. W., Li, L. X., \& Remillard, R. A. (2006). Estimating the spin of stellar-mass black holes by spectral fitting of the X-ray continuum. The Astrophysical Journal Letters, 636(2), L113-L116.
\bibitem{PC2020} Guo, M., \& Li, P. C. (2020). Innermost stable circular orbit and shadow of the 4D Einstein–Gauss–Bonnet black hole. The European Physical Journal C, 80(6), 588.
\bibitem{Bamba2016}Bamba, K. (2016). Thermodynamic properties of modified gravity theories. International Journal of Geometric Methods in Modern Physics, 13(06), 1630007
\bibitem{Clifton2012}Clifton, T., Ferreira, P. G., Padilla, A., \& Skordis, C. (2012). Modified gravity and cosmology. Physics reports, 513(1-3), 1-189.


\bibitem{kalb1974classical} Kalb, M., \& Ramond, P. (1974). Classical direct interstring action. Physical Review D, 9(8), 2273.

\bibitem{green1987superstring} Green, M. B., Schwarz, J. H., \& Witten, E. (2012). Superstring theory: volume 1, Introduction. Cambridge university press.

\bibitem{kao1996induced} Kao, W. F., Dai, W. B., Wang, S. Y., Chyi, T. K., \& Lin, S. Y. (1996). Induced Einstein-Kalb-Ramond theory and the black hole. Physical Review D, 53(4), 2244.

\bibitem{new1} Altschul, B., Bailey, Q. G., \& Kostelecký, V. A. (2010). Lorentz violation with an antisymmetric tensor. Physical Review D—Particles, Fields, Gravitation, and Cosmology, 81(6), 065028.

\bibitem{new2} Lessa, L. A., Silva, J. G., Maluf, R. V., \& Almeida, C. A. S. (2020). Modified black hole solution with a background Kalb–Ramond field. The European Physical Journal C, 80(4), 335.

\bibitem{new5} Duan, Z. Q., Zhao, J. Y., \& Yang, K. (2024). Electrically charged black holes in gravity with a background Kalb–Ramond field. The European Physical Journal C, 84(8), 798.

\bibitem{new3}  Yang, K., Chen, Y. Z., Duan, Z. Q., \& Zhao, J. Y. (2023). Static and spherically symmetric black holes in gravity with a background Kalb-Ramond field. Physical Review D, 108(12), 124004.

\bibitem{new4} Lessa, L. A., Silva, J. G., Maluf, R. V., \& Almeida, C. A. S. (2020). Modified black hole solution with a background Kalb–Ramond field. The European Physical Journal C, 80(4), 335.


\bibitem{new6} Wang, J., et al. (2026). Testing quantum-corrected black holes with Kalb–Ramond field effects via trajectories dynamics and frequencies spectra. International Journal of Geometric Methods in Modern Physics, 2650185.

\bibitem{new7} Atamurotov, F., et al. (2022). Particle dynamics and gravitational weak lensing around black hole in the Kalb-Ramond gravity. The European Physical Journal C, 82(8), 659.


\bibitem{KR}Sucu, E., \& Sakallı, I. (2025). Exploring Lorentz-violating effects of Kalb-Ramond field on charged black hole thermodynamics and photon dynamics. Physical Review D, 111(6), 064049.
\bibitem{Kalb-Ramond}Jumaniyozov, S., Murodov, S., Rayimbaev, J., Ibragimov, I., Madaminov, B., Urinbaev, S., \& Abdujabbarov, A. (2025). Black holes surrounded by PFDM in Kalb-Ramond gravity: from thermodynamics to QPO tests. The European Physical Journal C, 85(7), 797.

\bibitem{gravityjunior2024lensing} Junior, E. L. B., et al.  (2024) Gravitational lensing of a Schwarzschild-like black hole in Kalb-Ramond gravity,. , Phys. Rev. D 110, 024077

\bibitem{kumar2020rotating} Kumar, R., Ghosh, S. G., \& Wang, A. (2020). Gravitational deflection of light and shadow cast by rotating Kalb-Ramond black holes. Physical Review D, 101(10), 104001.

\bibitem{atamurotov2022particle} Atamurotov, F., Ortiqboev, D., Abdujabbarov, A., \& Mustafa, G. (2022). Particle dynamics and gravitational weak lensing around black hole in the Kalb-Ramond gravity. The European Physical Journal C, 82(8), 659.

\bibitem{jumaniyozov2025pfdm} Jumaniyozov, S., et al. (2025). Black holes surrounded by PFDM in Kalb-Ramond gravity: from thermodynamics to QPO tests. The European Physical Journal C, 85(7), 797.

\bibitem{kr1} Shi, Y. (2025). Influence of a Kalb-Ramond black hole on neutrino behavior. Journal of High Energy Physics, 2025(8), 1-27.

\bibitem{kr2}  Asrat, M. (2025). Kalb-Ramond field, black holes and black strings in (2+ 1) D. Journal of High Energy Physics, 2025(8), 1-53.

\bibitem{kr3}  Filho, A. A. (2025). Particle motion and thermal effects around a Kalb–Ramond black hole. The European Physical Journal C, 85(9), 1002.

\bibitem{kr4}  Sekhmani, Y., Maurya, S. K., Rayimbaev, J., Altanji, M., Ibragimov, I., \& Muminov, S. (2025). Lorentz-violating ModMax black holes in phantom-enhanced Kalb-Ramond gravity: Thermodynamics and topological charges. Physics of the Dark Universe, 102079.

\bibitem{Nojiri2011} Nojiri, S. I., \& Odintsov, S. D. (2011). Unified cosmic history in modified gravity: from F (R) theory to Lorentz non-invariant models. Physics Reports, 505(2-4), 59-144.

\bibitem{Capozziello2011}Capozziello, S., \& De Laurentis, M. (2011). Extended theories of gravity. Physics Reports, 509(4-5), 167-321.

\bibitem{Harko2014}Harko, T., \& Lobo, F. S. (2014). Generalized curvature-matter couplings in modified gravity. Galaxies, 2(3), 410-465.

\bibitem{e1} Chandrasekhar, S. (1998). The mathematical theory of black holes (Vol. 69). Oxford university press.
\bibitem{Alomar2025}Ditta, A., Bouzenada, A., Alomar, M., \& Bin-Asfour, M. (2025). Dynamics of test particles and quasi-periodic oscillations mathematically around F (R)-corrected black holes. Nuclear Physics B, 117099.
\bibitem{Ditta2025}Ashraf, A., Ditta, A., Bouzenada, A., Maurya, S. K., Abd-Elmonem, A., Suoliman, N. A., \& Channuie, P. (2025). Plasma lensing, epicyclic oscillations, particle collision, and thermal fluctuations around a short-hairy black hole. Physics of the Dark Universe, 48, 101836.
\bibitem{Bouzenada2024}Ditta, A., Bouzenada, A., Mustafa, G., Alanazi, Y. M., \& Mushtaq, F. (2024). Particle motion, shadows and thermodynamics of regular black hole in pure gravity. Physics of the Dark Universe, 46, 101573.

\bibitem{Saleem2025}Saleem, A., Ditta, A., Bouzenada, A., \& Alkahtani, B. S. (2025). Epicyclic oscillations, particle collisions, emission energy and thermal fluctuation around a black hole with de-sitter core. Nuclear Physics B, 1017, 116926.

\bibitem{e3} Charles W. Misner, Kip S. Thorne, and John Archibald Wheeler (1973): Gravitation. W. H. Freeman and Company, San Francisco.

\bibitem{Mustafa2025}Mustafa, G., Channuie, P., Javed, F., Bouzenada, A., Maurya, S. K., Cilli, A., \& Güdekli, E. (2025). Orbital motion and epicyclic oscillations around a black hole with magnetic charge. Physics of the Dark Universe, 47, 101765.\
\bibitem{Javed2025}Ashraf, A., Alqahtani, A. S., Javed, F., Channuie, P., Cilli, A., Bouzenada, A., ... \& Malik, M. Y. (2025). Orbital motion and QPOs testing around rotating Hairy black holes in Horndeski gravity. Physics of the Dark Universe, 47, 101725.
\bibitem{Bouzenada2025}Bouzenada, A., Ditta, A., Ashraf, A., Maurya, S. K., Atamurotov, F., Aslam, M., \& Malik, M. Y. (2025). Barrow entropy effects on thermodynamics and quasi-periodic oscillations around a Frolov black hole. Nuclear Physics B, 1017, 116928.
\bibitem{ISCO}Cen, Y., \& Song, Y. (2025). Upper bound on the radius of the innermost stable circular orbit of black holes. Physics Letters B, 866, 139545.
\bibitem{e2} Bardeen, J. M., Press, W. H., \& Teukolsky, S. A. (1972). Rotating black holes: locally nonrotating frames, energy extraction, and scalar synchrotron radiation. Astrophysical Journal, Vol. 178, pp. 347-370 (1972), 178, 347-370.
\bibitem{e4} Abramowicz, M. A., \& Fragile, P. C. (2013). Foundations of black hole accretion disk theory. Living Reviews in Relativity, 16(1), 1.

\bibitem{h1} Giulini, D. (2015). Luciano Rezzolla and Olindo Zanotti: Relativistic hydrodynamics: Oxford University Press, Oxford, 2013, 752 pp, GBP 55.00, ISBN: 978-0-19-852890-6.
\bibitem{h2} Kato, S. (2001). Basic properties of thin-disk oscillations. Publications of the Astronomical Society of Japan, 53(1), 1-24.
\bibitem{h3} Abramowicz, M. A., \& Kluźniak, W. (2001). A precise determination of black hole spin in GRO J1655-40. Astronomy \& Astrophysics, 374(3), L19-L20.
\bibitem{h4} Okazaki, A. T., Kato, S., \& Fukue, J. (1987). Global trapped oscillations of relativistic accretion disks. Publications of the Astronomical Society of Japan, 39(3), 457-473.

\bibitem{h5} Abramowicz, M. A., Karas, V., Kluzniak, W., Lee, W. H., \& Rebusco, P. (2003). Non-linear resonance in nearly geodesic motion in low-mass X-ray binaries. Publications of the Astronomical Society of Japan, 55(2), 467-471.
\bibitem{h6} Torok, G., Abramowicz, M. A., Kluzniak, W., \& Stuchlik, Z. (2005). The orbital resonance model for twin peak kHz quasi periodic oscillations in microquasars. Astronomy \& Astrophysics, 436(1), 1-8.
\bibitem{p1} Stella, L., \& Vietri, M. (1998). Lense-thirring precession and quasi-periodic oscillations in low-mass X-ray binaries. The Astrophysical Journal Letters, 492(1), L59-L62.

\bibitem{em3} Page, D. N. (1976). Particle emission rates from a black hole. II. Massless particles from a rotating hole. Physical Review D, 14(12), 3260.
\bibitem{em4} Wald, R. M. (1994). Quantum field theory in curved spacetime and black hole thermodynamics. University of Chicago press.
\bibitem{em1} Hawking, S. W. (1975). Particle creation by black holes. Communications in mathematical physics, 43(3), 199-220.
\bibitem{em2} Bekenstein, J. D. (1973). Black holes and entropy. Physical Review D, 7(8), 2333.

\bibitem{108}Övgün, A., Javed, W., \& Ali, R. (2018). Tunneling Glashow‐Weinberg‐Salam Model Particles from Black Hole Solutions in Rastall Theory. Advances in High Energy Physics, 2018(1), 3131620.
\bibitem{113}Ditta, A., Tiecheng, X., Ali, R., Atamurotov, F., Mahmood, A., \& Mumtaz, S. (2023). Thermodynamic stability of the regular charged torus-like black hole. Annals of Physics, 453, 169326.
\bibitem{114}Ditta, A., Tiecheng, X., Ali, R., \& Mustafa, G. (2023). Thermal stability and tunneling radiation in Van der Waals black hole. Nuclear Physics B, 994, 116287.

\end{thebibliography}
\end{document}